*Review*

# Review on Physically Flexible Nonvolatile Memory for Internet of Everything Electronics

**Mohamed T. Ghoneim and Muhammad M. Hussain \***

Integrated Nanotechnology Lab, Electrical Engineering, Computer Electrical Mathematical Science and Engineering Division, King Abdullah University of Science and Technology (KAUST), Thuwal 23955-6900, Saudi Arabia; E-Mail: mohamed.ghoneim@kaust.edu.sa

**\*** Author to whom correspondence should be addressed;
E-Mail: MuhammadMustafa.hussain@kaust.edu.sa; Tel.: +966-544-700-072.

**Abstract:** Solid-state memory is an essential component of the digital age. With advancements in healthcare technology and the Internet of Things (IoT), the demand for ultra-dense, ultra-low-power memory is increasing. In this review, we present a comprehensive perspective on the most notable approaches to the fabrication of physically flexible memory devices. With the future goal of replacing traditional mechanical hard disks with solid-state storage devices, a fully flexible electronic system will need two basic devices: transistors and nonvolatile memory. Transistors are used for logic operations and gating memory arrays, while nonvolatile memory (NVM) devices are required for storing information in the main memory and cache storage. Since the highest density of transistors and storage structures is manifested in memories, the focus of this review is flexible NVM. Flexible NVM components are discussed in terms of their functionality, performance metrics, and reliability aspects, all of which are critical components for NVM technology to be part of mainstream consumer electronics, IoT, and advanced healthcare devices. Finally, flexible NVMs are benchmarked and future prospects are provided.

## 1. Introduction

Recent advancements in flexible electronics research will enable novel applications ranging from stylish flexible gadgets for real-time monitoring of health-related vital signs to novel biological applications such as electronic skin [1–11]. Critical advances have been made in recent years that rely on organic materials as active elements because of their inherent flexibility. Mainstream approaches to capitalize on naturally flexible substrates like polymers can be categorized into (i) all-organic systems, where both devices (specifically active materials) and substrates are made up of organic materials [12–18] or inkjet- and screen-printed in thin layers onto paper and organic substrates [19–22] and (ii) hybrid systems, where inorganic electronic devices are transferred onto an organic substrate using transfer printing and other transfer techniques [23–31], laser lift-off transfer [32], and low-temperature direct deposition of inorganic devices on plastic organic substrates [33–36]. Other approaches use silicon-on-insulator (SOI) substrates, and controlled spalling technology to peel-off thin semiconductor layers [37–39]. In addition, a complementary transfer-free approach has recently been introduced, where thinning down the inorganic substrate through traditional, standard fabrication processes improves the flexibility of the substrate [40–45].

These approaches are all geared towards achieving fully flexible electronic systems. The three main components in any electronic system are (1) processing units; (2) the main memory; and (3) storage. Processing units perform logic operations through transistor logic, while the main memory performs temporary short-term storage (cache) with a quick access feature, also known as primary storage or random access memory (RAM), through access transistors and capacitors that store charges. Storage refers to the long-term retention of information, traditionally implemented using hard disks. However, at present, a shift towards other NVM types in a solid-state drive (SSD) format for supporting faster performance and higher integration densities within strict area constraints is taking place. Hence, the main electronic devices required to build an electronic system are transistors, capacitors, and NVM devices. Emerging NVM such as resistive random access memory (ReRAM), flash memory, phase change RAM (PCRAM), and ferroelectric random access memories (FeRAMs) have the benefits of fast switching, low-operation voltage, and ultra-large-scale-integration (ULSI) densities. These attractive qualities not only make them a favorable option for replacing magnetic hard disks but also for replacing quick access, volatile, dynamic RAM. This means that future electronic systems will require combinations of only two essential devices: transistors and NVM devices.

An objective assessment of the discussed mainstream and complementary approaches to flexible electronics must focus on their ability to provide high-performance, reliable NVM devices with ULSI density transistors. In this review, we present the mainstream NVM architectures and technologies with a special focus on most up-to-date techniques for producing flexible NVM devices.

Every memory cell consists of a gating device for access/select that is usually implemented using a transistor. Hence, memory arrays are where the largest number of transistors exist in an electronic system, a consequence of their ULSI density and low cost/bit ($/bit). Furthermore, with the continuous reduction in $/bit of NVMs and the higher switching speeds between memory states ('0' to '1' or *vice versa*) of emerging NVM technologies, replacing volatile random access memory and magnetic hard disks with faster SSDs made up of transistors and NVM structures becomes feasible. Here, the progress made over the past few years in three prominent types of flexible NVM technologies is discussed: (i) resistive; (ii) ferroelectric; (iii) phase change; and (iv) charge-trapping NVMs. In addition, the reliability aspects of the devices reported are discussed and an assessment for emerging technologies that provides useful insights towards their potential for commercialization is also provided. Figure 1 briefly positions the focus of this review in context with flexible electronics research.

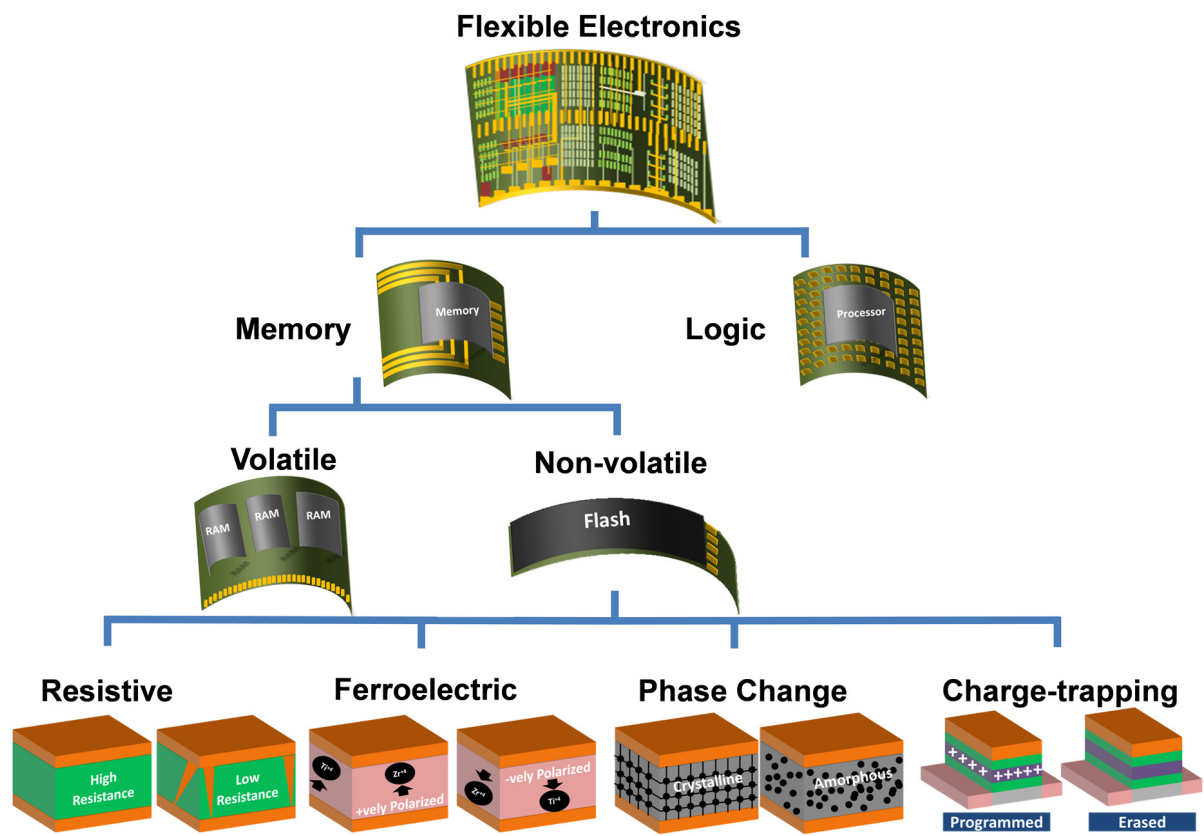

**Figure 1.** Chart highlighting the focus of the review.

In this review, our discussion is restricted to works that demonstrated an actual flexible version of flexible rewriteable NVMs between January 2010 and May 2015. However, there are many interesting works on flexible write-once-read-multiple (WORM) NVMs and other NVM devices that might be suitable for future applications in flexible electronics [46–51]. Furthermore, multiple interesting review papers are available on non-volatile memory technologies; however, the comprehensive scope of this review combining various flexing approaches, non-volatile memory and transistor technologies, arraying architectures with a special focus on flexibility, performance, reliability and monolithic integration ability has not been reported to date. For instance, Chung *et al.* [52], Makarov *et al.* [53], and Wang *et al.* [54] reviewed NVM devices but flexibility was not addressed. Naber *et al.* [55], Wang *et al.* [56], Liu *et al.* [57], and Chou *et al.* [58] reviewed only organic NVM. J.-S. Lee reviewed floating gate NVM devices [59]. Lee and Chen [60], Jeong *et al.* [61], Panda *et al.* [62], Lin *et al.* [63], Porro *et al.* [64], and Seok *et al.* [65] reviewed only resistive (ReRAM) NVM. Kim *et al.* reviewed hybrid organic/inorganic nanocomposites materials for NVM [66]. Liu *et al.* [67] and Mai *et al.* [68] reviewed only ferroelectric NVM and ferroelectric electronics. Kurosawa *et al.* reviewed polyamide based memory [69]. Han *et al.* reviewed various flexible NVM technology but reliability issues, flexible access and logic transistors, and monolithic integration ability were not addressed [70]. Acharyya *et al.* reviewed the reliability of TiO$_2$ based ReRAM [71], and Gale *et al.* reviewed only TiO$_2$ based ReRAM [72]. Therefore, this review is unique in its scope, providing a comprehensive perspective on the collective progress in the field of flexible electronics with a special focus on flexible nonvolatile memory technologies.

As of today variety of materials have been used to build NVM devices. For example, NMVs based on; (i) embedded 0-dimensional gold nanoparticles (NPs) [73–77], black phosphorous quantum dots (QDs) [78], and silicon QDs [79]; (ii) 1-dimensional zinc oxide (ZnO) nanowires [48], silicon (Si) nanowires [80], and carbon nanotubes (CNTs) [81–83]; and (iii) 2-dimensional graphene [49,84,85], graphene oxide [46,86–91], molybdenum disulfide (MoS$_2$) [50,92], zinc oxide (ZnO) [48], and hydrated tungsten tri-oxide (WO$_3$.H$_2$O) nano-sheet [29] have already been reported.

Although these and other similar reports have focused on discrete memory devices or cells, academic researchers and semiconductor industries have reported macroelectronics (large area electronics) focusing on mainly artificial skin [93–95] and display technology where memory has not been an integrated built-in module. Sony reported a rollable organic light emitting diode (OLED) based display in 2010 [96]. Samsung in the consumer electronics show (CES) 2011 and, later, LG and Nokia demonstrated a flexible display prototype [97]. However, the futuristic vision of the IoT where everything is connected, communicating, and making real-time decisions with artificial intelligence, with the associated emerging markets of big data analysis and machine-to-machine (M2M) technologies, would require more than flexible displays. The steep increase in the number of sensors from few millions in 2007 to billions of devices in 2013 is expected to persist, reaching the trillion sensor devices mark by 2023 due to the impact of IoT [98]. These sensors will be integrated in smart cards and RFIDs, vehicular technologies, buildings, infrastructures, healthcare, smart energy, factories, and supply chain management [98–100], as well as on humans for improving regular day to day experience [101]. To achieve such functionalities and fulfill the futuristic vision, IoT devices will require: (i) increased intra-node processing for real-time decision making; (ii) robustness to environmental variations (reliability); (iii) ultra-low power operation; (iv) ultra-high density integrated NVM and (v) smart antennas for wireless communications [102–106]. In addition, IoT devices should be physically flexible to enable wider adaptation in wearable forms and conforming to curvilinear structures in various forms.

To this end, integrated device manufacturers (IDMs) have already demonstrated devices designed to meet the IoT requirements. In 2014, Aitken *et al.* identified the 65 nm CMOS technology as the most suitable IoT chip process based on wafer cost and die area analysis [107]. In 2015, Yamamoto *et al.* demonstrated a novel gate stack in 65 nm CMOS technology for ultra-low leakage devices [103], Ueki *et al.* from Renesas Electronics Corporation developed a low power 2 Mb ReRAM macro in 90 nm CMOS technology [108], Whatmough *et al.* implemented a 0.6 volts transceiver in 65 nm CMOS technology [109], and Yamauchi *et al.* developed an embedded flash memory in vehicle control systems for IoT applications [100]. Furthermore, Hitachi researchers have studied how to profit from IoT for 10 years and used big data analysis to introduce the wearable happiness meter to unravel the link between physical motion and happiness [101]. Tanakamaru *et al.* introduced privacy protection solid state storage (PP-SSS) for what they called, "the right to be forgotten," where data is physically and automatically broken hardware-wise to co-op with anticipated security and privacy issues in the IoT era [110]. These are all great milestone in providing useful insights of what the future holds with the IoT revolution.

Based on the existing progress and current status, it is evident that while IoT devices are required to attain physical flexibility, they still have to rely on CMOS technology while pushing for ultra-low power consumption, ultra-low leakage currents, improved reliability, and ultra large scale co-integration of NVMs, CPUs, and antennas. Flexible antennas have been studied decades and will not be discussed in this review [111–116]. As aforementioned, NVM modules require information storage elements and select access transistors; therefore, a NVM perspective of the flexible electronics arena provides a comprehensive overview of the basic elements needed for implementing all electronic systems, including systems suitable for IoT applications.

## 2. Approaches for Making Flexible Devices

### 2.1. The All-Organic Approach

Most organic electronics use a variety of polymeric semiconductors as channel materials, polymeric ferroelectrics for nonvolatile storage, and thick, durable insulating polymers to support the flexible substrate. Figure 2a shows a representative structure for an all-organic deposited NVM that uses a quinoidal oligothiophene derivative (QQT(CN)$_4$) as the organic channel material and a polyvinilidene-co-trifluoroethylene (PVDF-TrFE) as the ferroelectric material; Figure 2b shows an inkjet-printed organic inverter on a plastic substrate. Compared to inorganic silicon electronics, the all-organic approach is more challenging with respect to the performance of organic materials, especially as transistor channel materials. The highest reported mobility for most organic channel materials is more than 20 times lower than silicon [117–120], with the exception of 43 cm$^2$/V.s peak hole saturation mobility reported by Yongbo Yuan *et al.* in 2014 [121]: this translates into lower performance. Furthermore, organic electronics still have to match the reliability of inorganic electronics nor can they compare in thermal stability [122]. To capitalize on the low-cost benefits of an all-organic system, there is a need to integrate polymeric dielectrics because they typically have low dielectric constants compared to the semiconductor industry's high-κ dielectrics and, in most cases, are even lower than that of SiO$_2$ [123]. Although it is a challenge to achieve the high capacitance values required for high-performance electronic devices using all-organic materials, there are also benefits from their use such as extremely high flexibility and conformal abilities. Currently, flexible organic electronic research has already gained solid grounds in commercial applications like active-matrix organic light-emitting diode (AMOLED) displays [124]. Therefore, organic electronics show true potential for further expansion and enhanced maturity in macroelectronics.

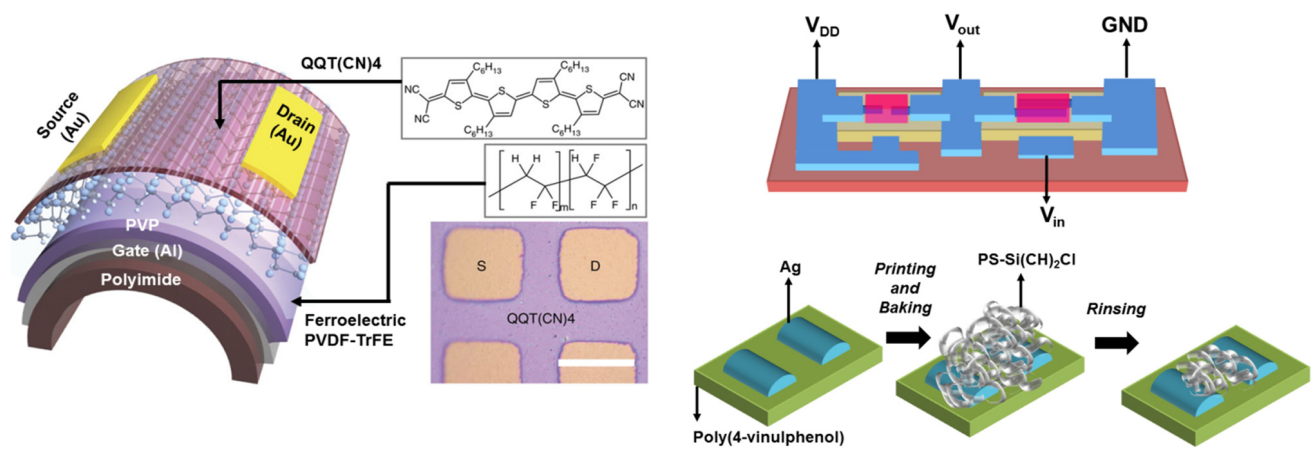

**Figure 2.** (**a**) Schematic representation of the devices and molecular structure of the organic semiconductor quinoidal oligothiophene derivative (QQT(CN)4) and the ferroelectric polyvinilidene-co-trifluoroethylene (PVDF-TrFE). Reprinted by permission from Macmillan Publishers Ltd.: Nature Communications [13], copyright (2014); (**b**) schematic of an all-inkjet-printed inverter using two p-type OTFTs (top) and diagrams of the PS brush treatment procedure on the PVP gate dielectric and Ag S/D electrodes (bottom). Reprinted with permission from [21]. Copyright © 2013 WILEY-VCH Verlag GmbH & Co. KGaA, Weinheim, Germany.

*2.2. The Hybrid Systems Approach*

Hybrid systems use both organic and inorganic materials, making them conducive to a wider spectrum of techniques with greater versatility. Figure 3 is an illustrative summary of available flexible hybrid techniques. In transfer printing, a molded polymer is used as a stamp that can be functionalized with desired materials and then printed onto a different substrate. Figure 3a shows the three modes of transfer printing by John Rogers's group at the University of Illinois at Urbana-Champaign [23]. Figure 3b shows a representative generic transfer technique, where devices are fabricated on a specific rigid substrate and then transferred to one that is flexible (a banknote in this case). A specific type of transfer is the laser lift-off by Keon Jae Lee's group at the Korea Advanced Institute of Science and Technology (KAIST), where a laser shot is used to etch a sacrificial layer to release the device for transfer (Figure 3c) [32]. Figure 3d shows a resistive memory structure made up of room-temperature-sputtered and e-beam-evaporated materials on a flexible substrate [33]. The mainstream hybrid transfer approach achieves high performance by transferring high-performance inorganic devices onto an organic substrate for flexibility; however, it adds extra nonconventional transfer steps and suffers low yields. Although this is mitigated by the direct-deposition-on-plastic-substrates approach, using plastic adds temperature restrictions to the fabrication process. As a result, it is a challenge to produce high-quality films such as atomic-layer-deposited high-κ dielectrics that usually require temperatures above 300 °C. Finally, because the different solvents used for patterning and photolithography should not affect the flexible organic substrate, there are limitations to suitable plastic material choices.

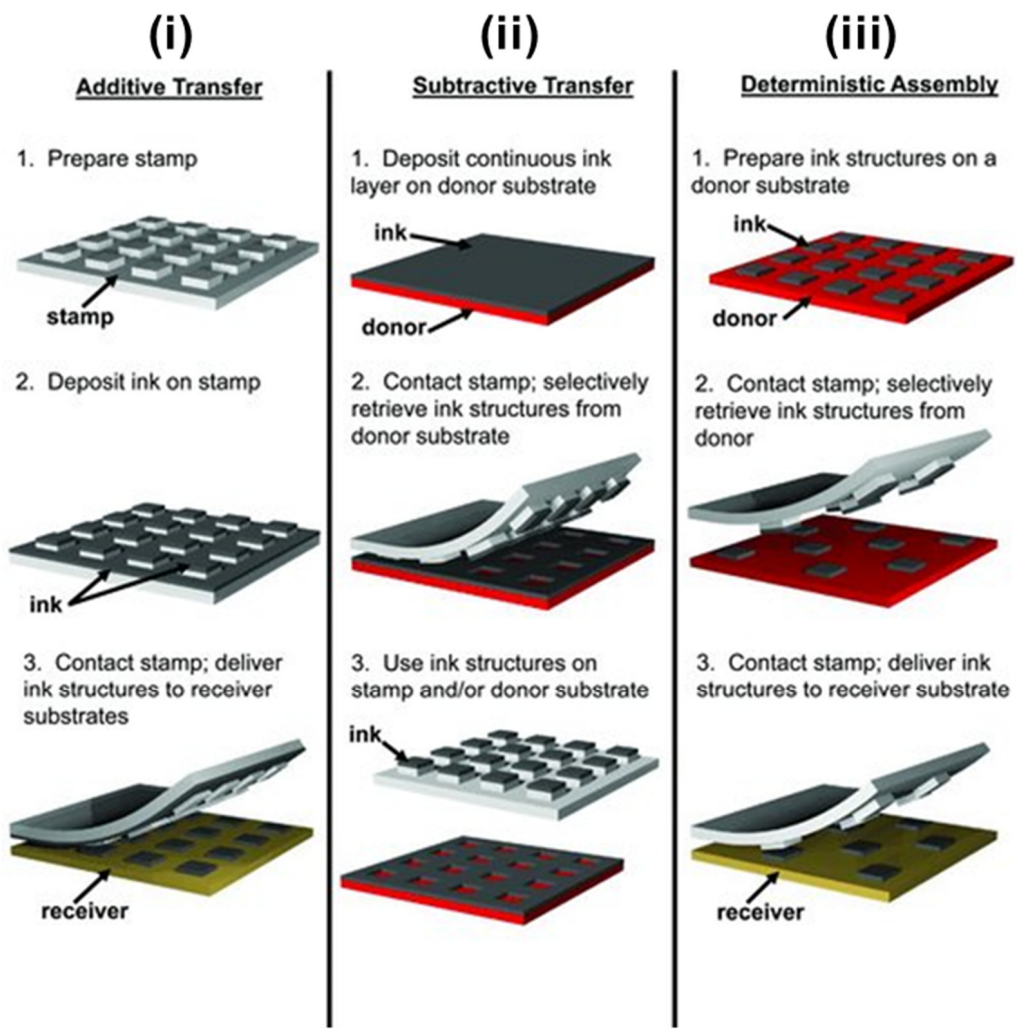

(**a**)

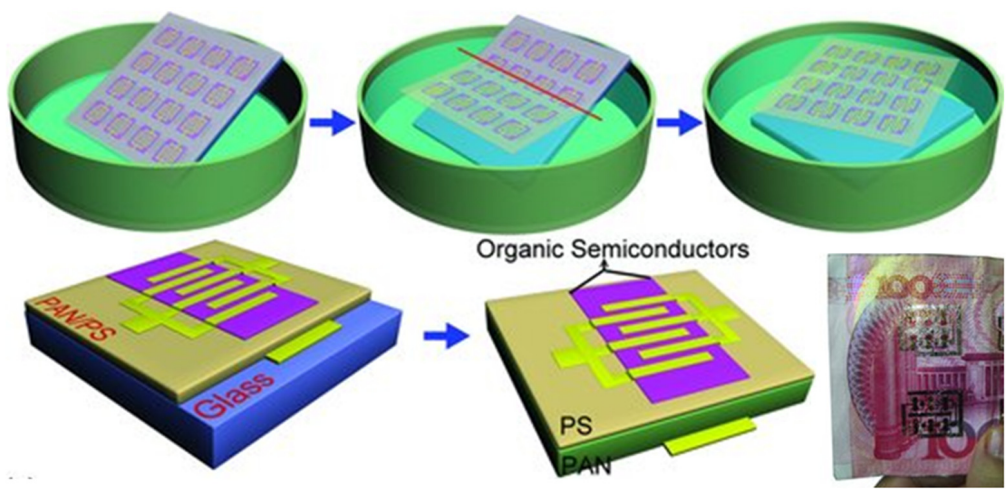

(**b**)

**Figure 3.** *Cont.*

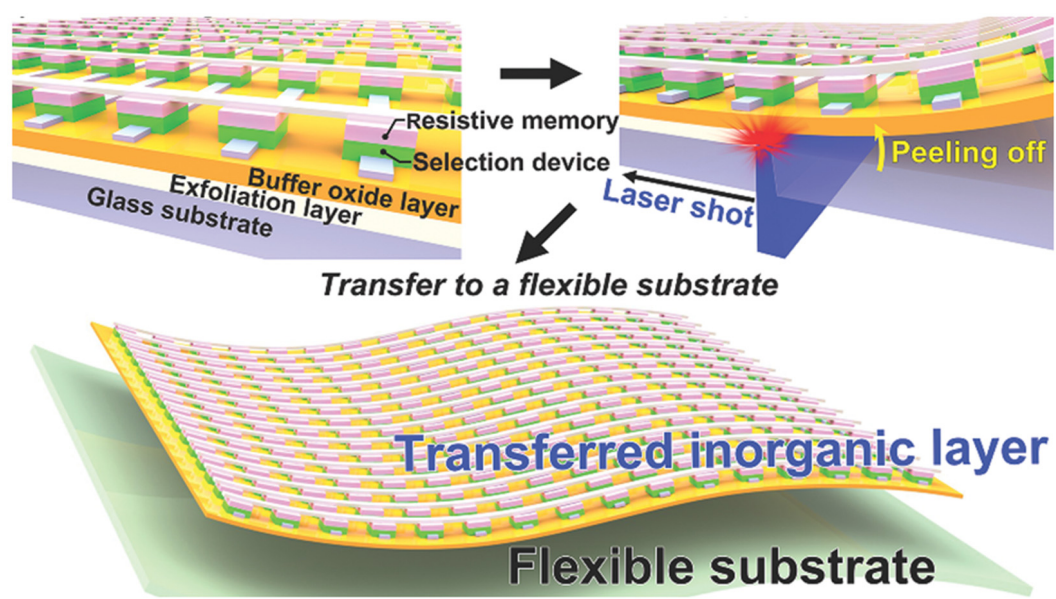

(**c**)

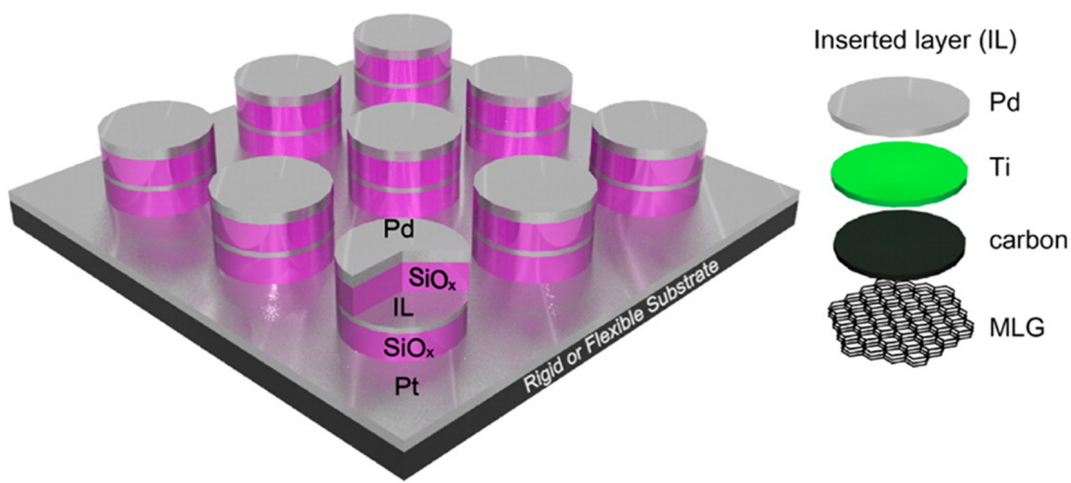

(**d**)

**Figure 3.** (**a**) Schematic illustrations of three basic modes for transfer printing. Reprinted with permission from [23]. Copyright © 2012 WILEY-VCH Verlag GmbH & Co.; (**b**) schematic diagram of the fabrication procedures for the freestanding OFETs using modified water-floatation method. Reprinted with permission from [30]. Copyright © 2013 WILEY-VCH Verlag GmbH & Co.; (**c**) schematic illustrations of the process for fabricating flexible crossbar-structured memory on a plastic substrate via the laser lift-off transfer method. Reprinted with permission from [32]. Copyright © 2014 WILEY-VCH Verlag GmbH & Co.; (**d**) schematic illustration of the cells in the conducting-interlayer SiOx memory device sputtered at room temperature. Reprinted with permission from [33]. Copyright © 2014 WILEY-VCH Verlag GmbH & Co.

*2.3. Spalling Technology*

The spalling technique uses stressor layers to initiate fracture-modes in SOI and semiconductor substrates. In 2012, Banarjee *et al.* reported on the exfoliation of thin-film transistors from prefabricated bulk wafers using the spalling technique [125]. In the same year, Shahrejerdi *et al.* reported having performed, at room temperature, controlled spalling of full SOI wafer circuitry and successfully transferred the surface layer to a flexible plastic substrate [37,126]. The authors deposited a nickel (Ni) stressor layer that abruptly discontinued near one edge of the wafer where a crack in the mono-crystalline silicon (Si) was initiated by an applied force [127,128]. However, before the force is applied, polyimide tape is added to support the flexible peeled fabric-bearing ultra-thin body devices (Figure 4). This approach has also been reported for single crystals of germanium (Ge) and gallium arsenide (GaAs) [38,129]. Another spalling approach was used by Bellanger and Serra where Si (100) foils were peeled off from the bulk substrate [39]. The challenges faced by the spalling technique are two fold: first, extra deposition and complex tuning of a stressor material with a specific thickness followed by etching are required and second, once the crack has been initiated, the peeling-off process requires high dexterity that is not suitable for mass production.

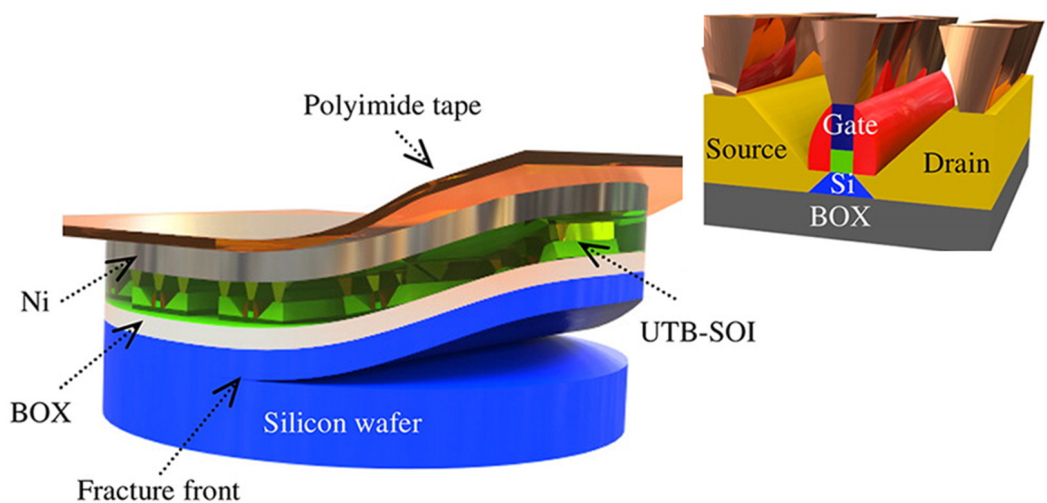

**Figure 4.** Schematic illustration of the controlled spalling process used for removing the prefabricated devices and circuits from the rigid silicon handle wafer. The inset schematically shows the device architecture for the ultra-thin body transistors with raised source/drain regions. Reprinted with permission from [37]. Copyright (2012) American Chemical Society.

*2.4. The Complementary Transfer-free Inorganic Approach*

The complementary transfer-free approach uses a fundamental inverse proportionality between the material's thickness and flexibility (Equation 1) to transform traditional, rigid electronic devices on economical Si (100) wafers into new, flexible devices by sufficiently reducing the thickness of the host's substrate.

$$[E \; \alpha \; \frac{1}{t^3}] \tag{1}$$

This approach provides a pragmatic solution to the aforementioned critical challenges by copying the associated perks of high performance, reliability, ULSI density, and the low cost of inorganic silicon-based electronics to the flexible arena via a transformed version of traditional devices. Moreover, the silicon industry has capitalized on monolithic integration over the past few decades and because of its core competitive advantages, it has grown into a huge market. Hence, preserving monolithic integration by using Si as a flexible substrate further improves this flexing approach. Figure 5 lists the different silicon-flexing techniques [40,44,45]. The scanning electron microscope image in Figure 5b illustrates the added advantage of extra device active area, which may be available in the form of conformal deposition of device layers through release trenches (holes) [130]. The etch-protect-release approach incurs some lost area, which is compensated by the potential reliability associated with the relatively novel air-gap shallow trench isolation (STI) technology [131]. On the other hand, remaining portions of the wafer can be recycled after chemical mechanical polishing and the holes network has a self-cooling effect, acting as air cooling channels for heat dissipation [132].

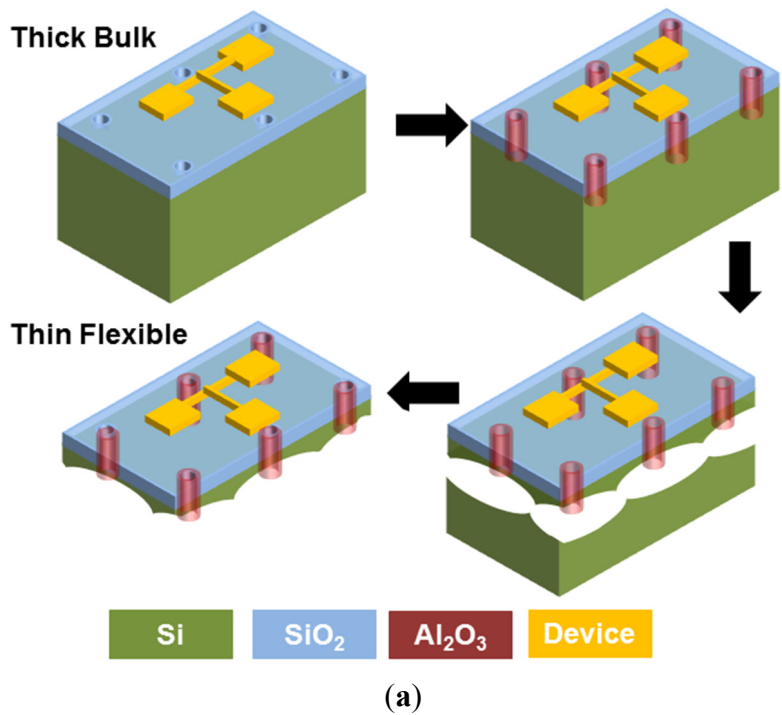

(a)

**Figure 5.** *Cont.*

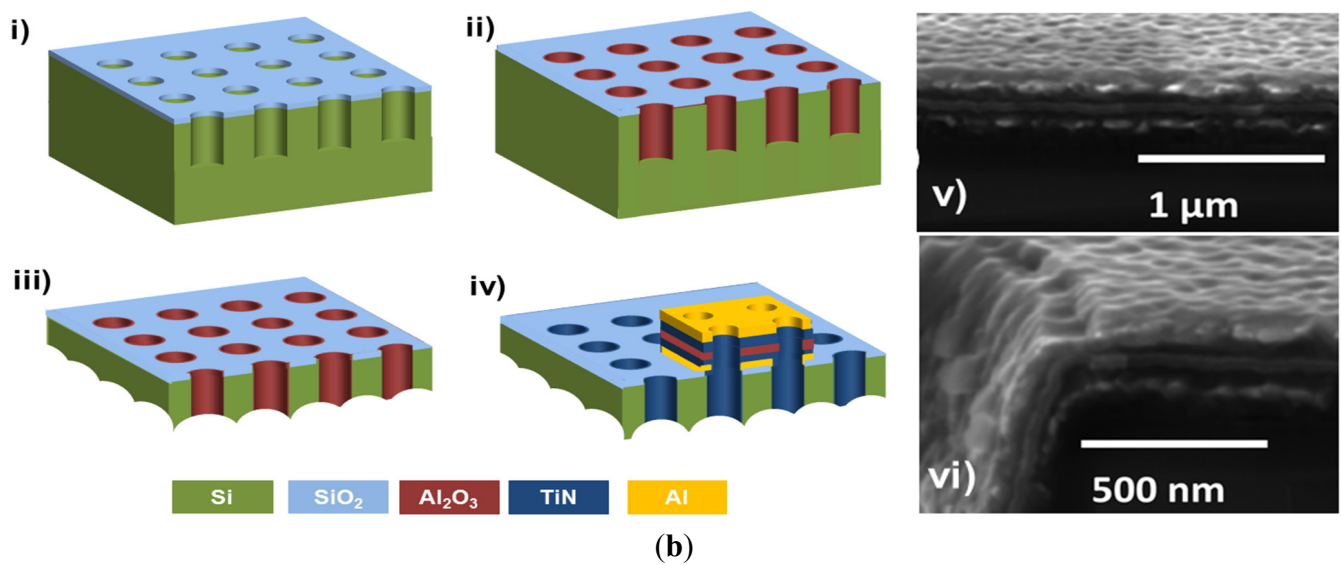

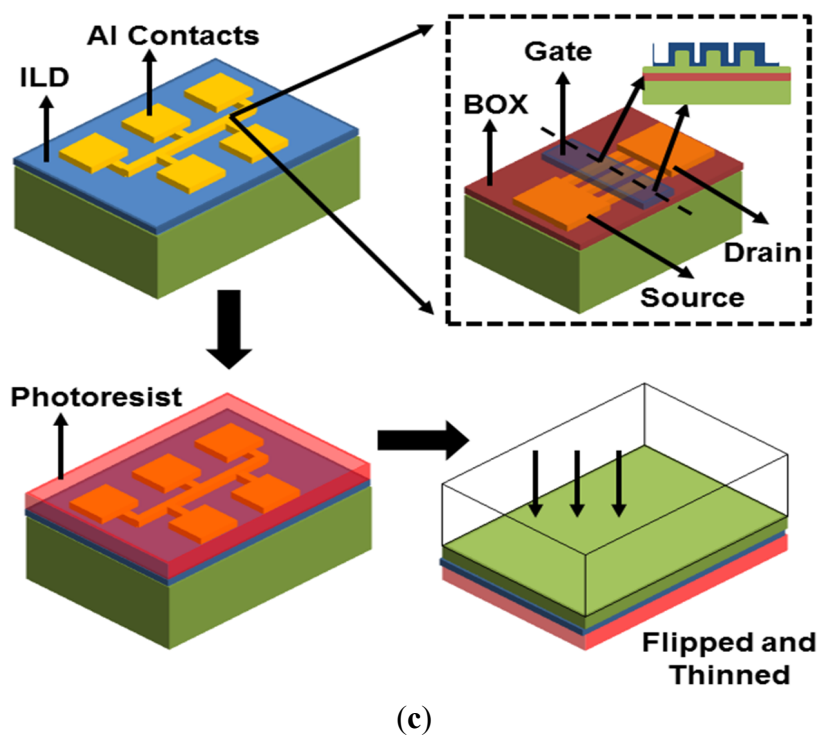

**Figure 5.** (**a**) Device first approach illustration where the devices are fabricated in a traditional fashion, then, protected using photoresist (PR). The PR is then patterned and the pattern is transferred to the field oxide (FOX) layer then to the Si substrate. Using the spacer technique, a highly conformal atomic layer deposition (ALD) spacer is deposited for sidewalls. Finally, the dies are put in a reactive chamber containing XeF2 gas for isotropic Si when the etching regions merge, the top flexible portion of Si (100) containing the devices can be safely released; (**b**) device last approach illustration where flexible silicon fabric is first released then devices are built. Adapted with permission from [130]. Copyright © 2014 WILEY-VCH Verlag GmbH & Co.; (**c**) illustration of the soft-etch back approach where the traditional dies containing devices are covered with PR for protection, then, the die is flipped upside down and etched using DRIE to the desired thickness. Adapted with permission from [45]. Copyright (2014) American Chemical Society.

Although conceptually the soft back-etching process (Figure 5c) is similar to the traditional back-grinding technique, there are considerable differences. For example, the soft back etch is a simple and delicate process compared to the complex and abrasive nature associated with the induced scratches, crystal defects, and the formation of stresses that take place during back grinding [133]. Furthermore, using the soft back etch requires no chemical mechanical polishing and leaves no residual stress on the substrate unlike the machining stress caused during back grinding [134–136]. Finally, back grinding to thinner substrates causes subsurface damage [136] and shallow surface cracks [135].

Fracture strength is a property that applies to all techniques because it determines the overall mechanical stability of an ultra-thin flexible electronic system [137]. To assess the fracture strength of a substrate, the most common method is the three-point bending test [138]. For Si thicknesses greater than 100 μm, the linear elastic bending beam theory provides an accurate estimation of fracture strength [139,140]; however, thinner substrates produce a nonlinear deflection-load relationship that is used to estimate fracture strength (Figure 6). To account for this nonlinearity, in 2015 Liu *et al.* introduced the large deflection theory of beam [138]. This relationship provides important insights for theoretical limitations of flexible silicon thinner than 100 μm. Furthermore, based on the application's required bending radius, the thickness of the flexible silicon substrate must be adjusted such that the applied stress ($stress = Young's\ modulus\ x\ strain\ (\varepsilon)$, where nominal strain is defined as $\varepsilon_{nominal} = thickness\ (t)/(2\ x\ bending\ radius\ (r))$) is lower than the fracture stress (fracture stress of thin Si substrates is higher than that of thicker substrates). For instance, [138] shows that for a 50-μm thick silicon substrate, the fracture stress is ~ 1.1 GPa. At the lowest estimate for Si (100), the Young's modulus is 128 GPa [141]; hence, the minimum bending radius that would cause fracture stress for a 50-μm thick flexible silicon substrate is ~ 3 mm and decreases with decreasing thicknesses. Therefore, a vanilla flexible silicon substrate that is 50-μm thick or less with a bending radius of > 3 mm will safely operate below the fracture stress level. We would like to point out that these results are for a bare silicon (100) substrate with no additive layers or patterns for devices. Therefore, based on the properties of the material, thickness of the substrate, and the bending radius necessary for a specific application, the most suitable approach and material system can be determined.

As identified by the International Technology Roadmap for Semiconductors (ITRS) 2013 Emerging Research Devices report, the main challenge will be to identify replacement technologies for static RAM and flash as they approach their physical limits [142]. Replacements must provide electrically accessible, high-speed, high-density, low-power NVMs that meet the reliability requirements for the desired devices including surviving high temperatures. It will be important to identify and address other reliability issues early in the development process. The temperature requirement for flexible, inorganic silicon-based NVM devices is a given because the materials used will have already survived the high thermal budgets required for the deposition of high-quality thin films used in complementary metal oxide semiconductor (CMOS) technology and subsequent front-end-of-line processing anneals. Due to the advancements in lithography, properties of a flexible inorganic NVM also rely on high integration density.

Challenges common to both organic and inorganic material systems toward their application in future electronics, including IoT devices include (i) attaining high speeds; (ii) being suitable for low-power devices; and (iii) identifying and assessing the reliability issues that arise when devices are flexed beyond the standard studied stresses in planar substrates. In addition, organic electronics must also achieve temperature stability and satisfactory integration density.

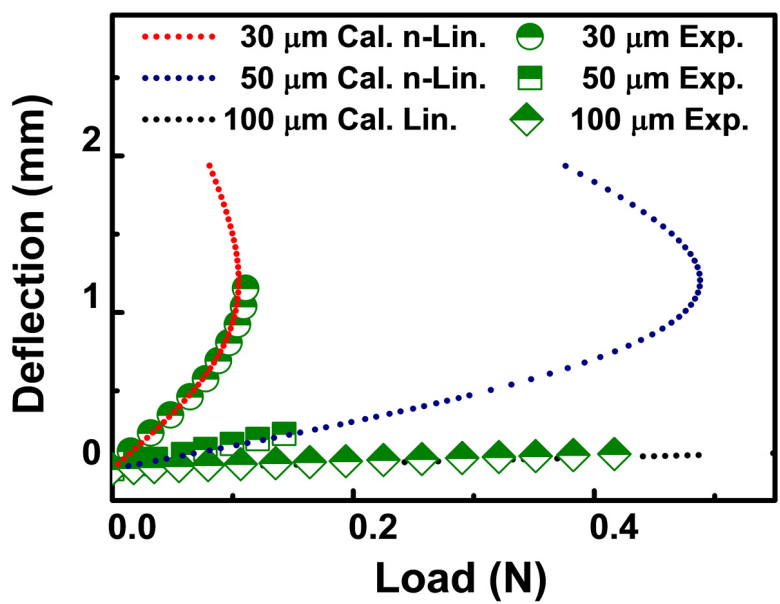

**Figure 6.** Deflection versus applied load plot for various thicknesses of flexible substrates, dotted lines showing non-linear analytical solution for 30 and 40 μm thick substrates and linear analytical solution for 100 μm thick substrates. Adapted courtesy of Prof. YongAn Huang, Huazhong University of Science and Technology, China.

## 3. NVM Operational Principles and Architectures

### 3.1. NVM Operational Principles

Similar to how we define the human brain's ability to memorize as the ability to remember useful information over long- and short-term durations, electronic memories have the ability to retain stored information over various durations. An electronic memory that is able to retain information over short periods of time (milliseconds) is identified as a volatile memory. In this case, when the power goes off, information stored in the volatile memory is lost. On the contrary, an electronic memory that is able to store information over long periods of time (~10 years is the industry standard) is called a nonvolatile memory (NVM). NVMs can retain information even when no power is supplied. There are five major classes of NVMs [70]: resistive RAM (ReRAM) also referred to as memristor [143–146], ferroelectric RAM (FeRAM), [147] magnetic RAM (MRAM) [148,149], phase change RAM (PCRAM) [150,151], and flash memory (floating gate (FG) and charge trapping (CT)) [80,152,153]. Other technologies, such as nano-electromechanical (NEM) NVMs [154,155] and molecular based NVMs [156] exist but they are not mainstream. Table 1 summarizes the principles of operation of the leading NVM technologies and indicates which technologies have already been demonstrated in a flexible form. Note that the terms 'floating gate' and 'charge-trapping flash' are used interchangeably in recent literature. In Table 1, the distinguishing property is whether the charge-trapping layer is a conductor or an insulator, although both conducting and insulating layers (with or without embedded NPs and QDs) trap charges; nanoparticles (NP) embedded in an insulator for charge trapping are also considered FG-Flash.

**Table 1.** Summary of non-volatile memory technologies and indications of flexed types.

| | NVM Type | Operation Principle | Flexed |
|---|---|---|---|
| 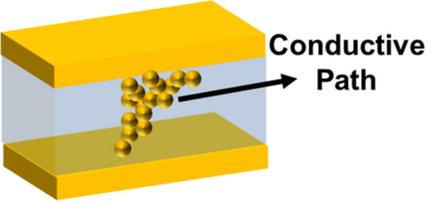 | ReRAM (Memristor) | A resistive oxide is sandwiched between two metallic layers. The resistance of the oxide changes with applied "set" and "reset" voltage pulses. A high-resistance state corresponds to "0" and a low-resistance state corresponds to "1". | Yes |
| 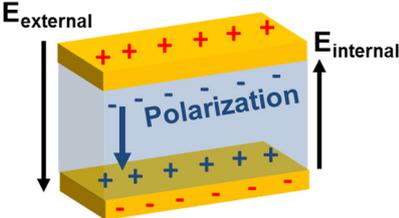 | FeRAM | A ferroelectric material has two possible polarization states inherent from its crystalline structure. Applying write/erase voltage pulse switches for positive to negative polarization states, corresponding to "0" or "1". | Yes |
| 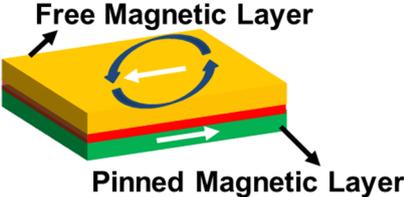 | MRAM | Spintronic devices such as magnetic tunneling junctions are composed of a fixed ("pinned") magnetic moment layer, a tunneling barrier (oxide), and a free layer. Current flowing in nearby lines is expected to magnetize the free layer. If the free layer magnetic moment is parallel to that of the pinned layer, the device is "ON" and the resistance across the structure is low. If the free layer is magnetized such that its magnetic moment is anti-parallel to the pinned layer, the device is "OFF" and the structure will be in high-resistance state. | No |

**Table 1.** *Cont.*

| | NVM Type | | Operation Principle | Flexed |
|---|---|---|---|---|
| 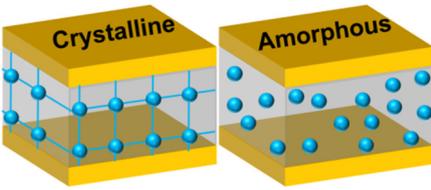 | PCRAM | | Current or laser pulses are applied to change the phase of a material from crystalline (low resistance) to amorphous (high resistance) and vice versa at a localized space, which changes the material's electrical and optical properties. Short pulses above the melting temperature are needed to make the change from the crystalline to the amorphous phase, while longer pulses below the melting temperature are required to restore the crystalline order of the material. | Yes |
| 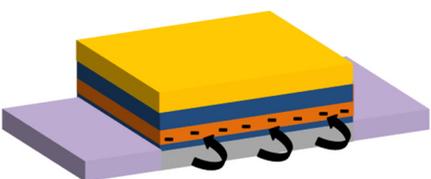 | Flash | FG | FG flash has the same structure as a field effect transistor (FET) except that its gate dielectric is split into three layers. The first is tunneling oxide, the second is an embedded conductor layer (*i.e.*, doped polysilicon or embedded quantum dots (QDs) or metallic nanoparticles (NPs)) floating gate, and the third is a blocking oxide. When a programming voltage is applied, carriers tunnel from the channel to the floating gate. This results in a shift of the threshold voltage of the transistor corresponding to "1". A reverse bias is applied during the erase operation to move the charges back into the channel. | Yes |

Table 1. *Cont.*

| NVM Type | Operation Principle | Flexed |
|---|---|---|
| CT | The charge trap flash replaces the floating gate with a conductor layer that has an insulting layer (*i.e.*, silicon nitride). The most common structures are the SONOS (Polysilicon-oxide-nitride-oxide-silicon) and the TANOS (titanium-alumina-nitride-oxide-silicon). | Yes |
| 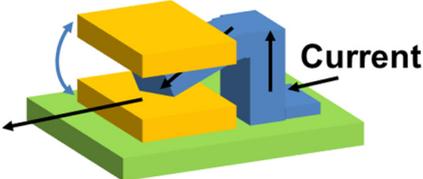 NEM-NVM | A nano-electromechanical switch is fabricated such that (i) upon applying a programming electrical signal, its pull-in voltage shifts when operated at a designed switching voltage or (ii) it has a free moving cantilever that has bistable physical states affecting its electrical properties. | No |
| 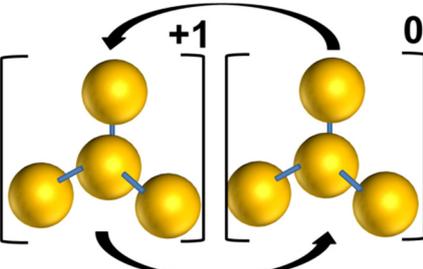 Molecular Based NVM | A bistable molecule can be switched from a low-conductance state ("0") to a high-conductance state "1" by applying brief bias voltage pulses to switch the state through oxidation and translation of the molecular structure between the two stable states. | No |

*3.2. NVM Architectures*

NVM architectures are an important element in memory design that can be classified into three main categories: the 1T, where the memory cell is composed of a single transistor ('T' stands for transistor); the 1T1C or 1T1R, where the memory cell is composed of an access/select transistor and a nonvolatile storage structure ('C' stands for capacitor and 'R' stands for resistor); and the 2T2C (two transistors and two capacitors per memory bit) [70,157,158]. Other variations of these main architectures [157,159] and different arrangements, such as the 1T2C, have also been reported [160–162]. Furthermore, there are differences in the way memory cells are connected to each other. For instance, NOR-type flash and NAND-type flash memories both have a 1T architecture but different cell connections [163]. Also, there is the crossbars configuration in which each memory cell is connected to four neighboring cells [145]. Figure 7 shows the schematic arrangements of the three main architectures, NOR and NAND flash arrangements, and the crossbars configuration.

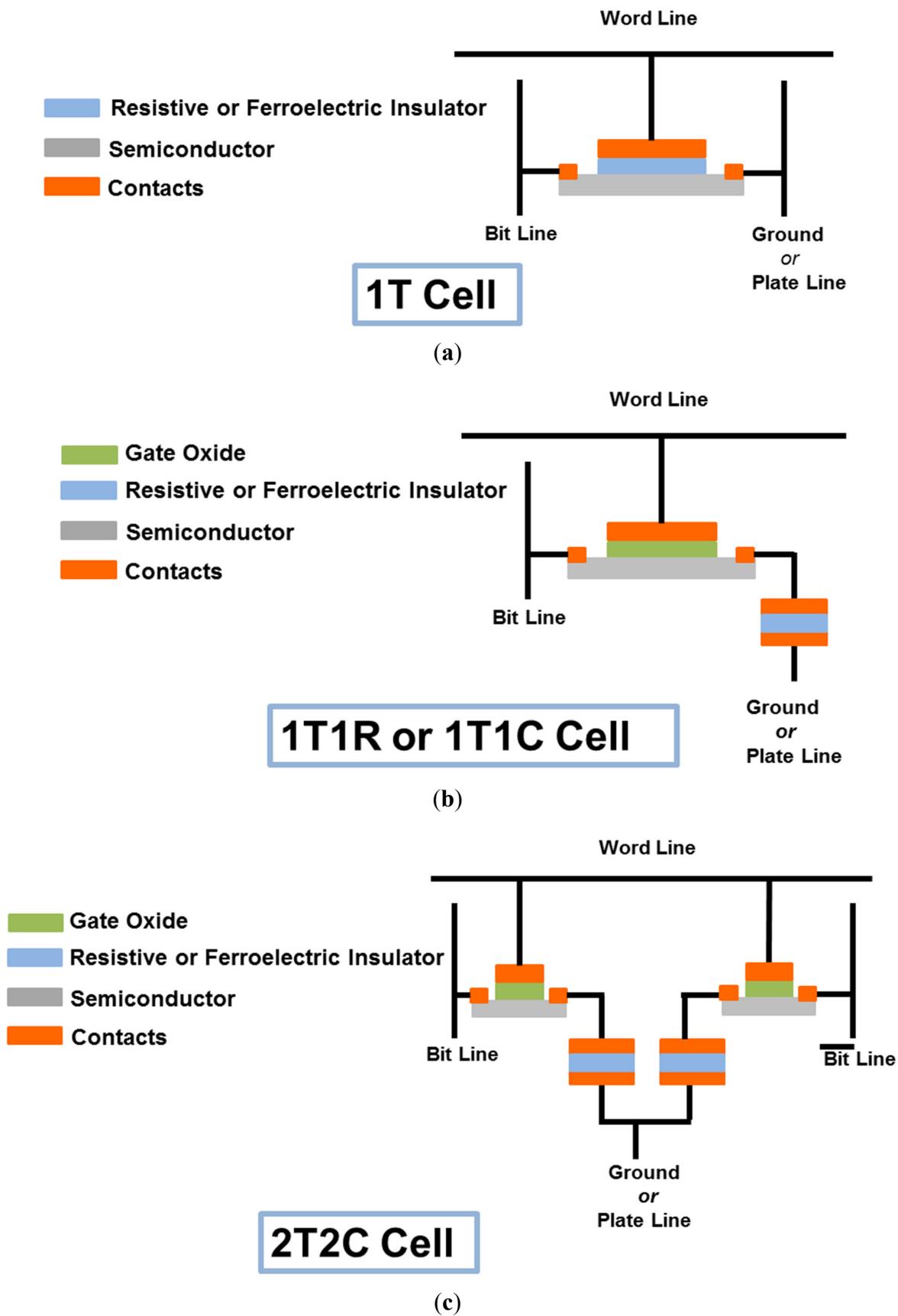

**Figure 7.** *Cont.*

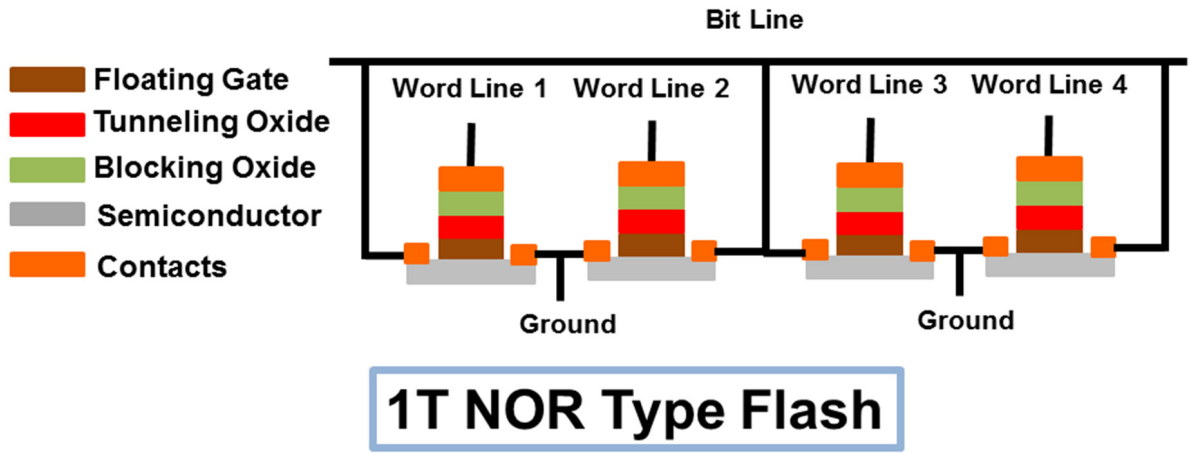

(**d**)

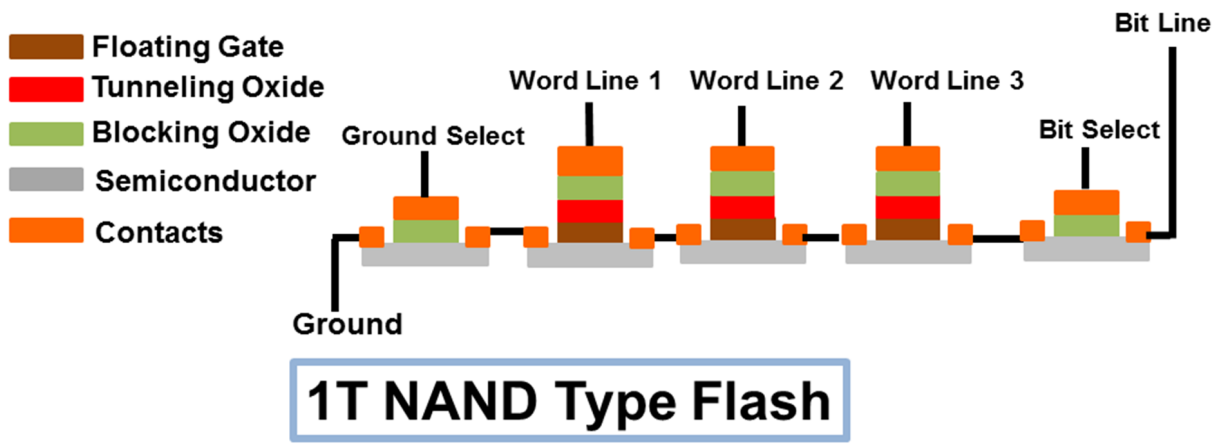

(**e**)

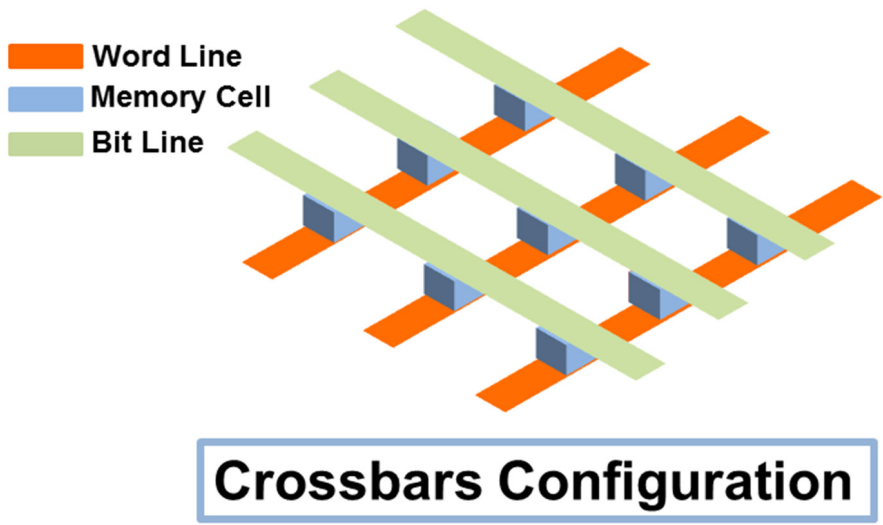

(**f**)

**Figure 7.** (**a**)–(**f**) Common memory architectures.

## 4. Flexible Field-Effect Transistors (FETs) for Logic and NVM Arrays

Today's memories have the highest density and number of integrated transistors of all electronic systems. With the advent of big data, the introduction of cloud computing, and the extensive deployment of gigantic data centers, the demand for more storage space and high storage density is increasing. Therefore, it has become essential to survey state-of-the-art flexible transistors. We think it is important to note that these transistors are not to be confused with the transistors used in 1T memory architectures. Although both are field-effect transistors, their functionality is conceptually different. One function is to store/trap charges while the other is to act as an electronic switch. 1T transistors are evaluated based on their ability to store charges (retention), memory endurance, and their ability to shift the voltage threshold (memory window) properties, while access/select and logic transistors are evaluated with respect to their ability not to store charges, subthreshold slopes, drain-induced barrier lowering, and stability of the threshold voltage. In this section, we focus on the switching-type transistor that is used in circuit logic and as a select/access transistor for the storage structures in 1T1C, 1T1R, and 2T2C architectures.

Over the past half-decade, a large portion of the reported work on flexible transistors has focused on thin film transistors (TFTs), which are a subset of FETs. TFTs are common in display technology for controlling active matrix organic light emitting diode (AMOLED) pixels [164]. Recently, however, they became popular for switching and logic computations due to their facile low-temperature fabrication on flexible substrates. The next subsections summarize the demonstrated flexible organic, hybrid, and inorganic FETs (mostly TFTs) over the past five years.

*4.1. All-Organic Transistors*

Table 2 highlights the demonstrated flexible all-organic transistors from 2010 up to May 2015. The best reported values are highlighted throughout for each performance metric; however, this does not mean that the most up-to-date technology is capable of being combined into one flexible all-organic structure with all the best performance because in most cases values are intertwined such that optimizing one value affects the others. For instance, increasing the thickness of the $Ba_{0.7}Sr_{0.3}TiO_3$ (BST) gate dielectric in an organic TFT can reduce the threshold voltage ($V_{th}$), increase the effective mobility ($\mu_{eff}$); on the other hand, it can reduce the $I_{ON}/I_{OFF}$, and increase the subthreshold slope (SS) [165]. In general, evidence suggests that the all-organic transistors suffer low mobilities (<10 $cm^2$/V.s), with the exception of graphene TFT, which achieves 340 $cm^2$/V.s [166]. We would also like to highlight the excellent flexibility of the devices down to a bending radius of 100 μm [167], and persistence up to 10,000 bending cycles [168]. The minimum feature reported is in the order of tens of micrometers (with the exception of Min *et al.* nanowires arrays, in 2013, achieving sub-micron dimensions [120]), which is relatively large as is the operational voltage and SS in many of the demonstrated devices. The highest reported yield for all organic transistors is 66%, reported for ink-jet printed devices [169].

*Electronics* **2015**, *4* 443Table 2. Key works on flexible all-organic transistors from 2010 up to May 2015. Highlights showing the best values reported.

| Reference | [165] | [170] | [171] | [167] | [169] | [172] | [173] |
|---|---|---|---|---|---|---|---|
| Year | 2012 | 2012 | 2012 | 2010 | 2013 | 2013 | 2013 |
| Structure | PEN/Ag gate BST dielectric/pentacene channel/Ag source and drain | PET/Al source and drain/C$_{60}$ channel/Parylene-C dielectric/Al gate | Metal gate/Mylar/TIPS-pentacene/Metal source and drain | PI/Al gate/ hybrid (AlO$_x$)-organic self-assembled monolayer (SAM) gate dielectric/Pentacene p-channel/Au source and drain | PET/ SWCNT/Ag source and drain/Barium titanate NP in PMMA ink dielectric/Ag gate | Au gate/Parylene dielectric/Au source and drain/TTC18-TTF channel | PI/gate electrode/Parylene dielectric/C$_8$-BTBT channel/Au source and drain/Parylene/PI |
| Approach | All organic | All organic | All organic | All organic | All organic (Ink-jet Printed) | All organic | All organic (Ink-jet Printing) |
| Dimensions (µm) | 100 × 2000 | 70 × 1800 |  | 50 × 500 | 85 × 1250 | 20 × 5000 | length is 20 |
| Mobility (cm$^2$/VS) | 0.53–1.24 | 0.58 | 0.1 to 0.4 | 0.5 | 9 | 0.0043 | 0.15 |
| SS (mV/Dec) | 100–160 | 1250 | 5000 | 166 |  |  | ~3300 |
| V$_{th}$ (V) | −1.11 to −1.18 | −0.1 | ~10 to 30 | ~−0.5 | 1.15 | 20 | ~10 |
| Operation Voltage (V) | 3 |  | 10 to 40 | 3 | −10 to 10 |  | 10 to −50 |
| min bending radius (mm) | 3 | 5.1 | 0.26 unencapsulated and 0.11 encapsulated in Parylene | 0.1 | 1 | 2 | 1 |
| I$_{ON}$/I$_{OFF}$ | 10$^4$ | 10$^5$ | 5 × 10$^3$ | 1.5 × 10$^3$ | 10$^5$ | 500 | 5 × 10$^4$ |
| Yield |  |  |  |  | 66% |  |  |
| Bending cycles |  |  |  |  |  |  |  |

**Table 2.** *Cont*.

| Reference | [168] | [174] | [175] | [176] | [166] | [120] | [177] |
|---|---|---|---|---|---|---|---|
| Year | 2014 | 2014 | 2015 | 2010 | 2014 | 2013 | 2015 |
| Structure | PI/Al gate/c-PVP/c-PVP/Pentacene/Au source and drain | PET/Ag source and drain/barium titanate-BTO-poly(methyl methacrylate) dielectric/Ag gate | PET/ITO gate/PVP dielectric/Pentacene channel/Au source and drain | PEN/Al gate/PI dielectric/QQT(CN)$_4$/Cr-Au source and drain | PET/Ti-Au source and drain/Graphene/PMMA/Au gate | PAR/Au source and drain/P3HT:PEO-blend NW/ion-gel polyelectrolyte/Au gate | PEN/PVP/Ag gate/Parylene dielectric/Ag source and drain/DTBDT-C$_{60}$ semiconductor |
| Approach | All Organic | All Organic (Screen Printing) | All Organic | All Organic | All Organic (Low Temperature Deposition + Graphene Transfer) | All organic | All organic |
| Dimensions (μm) | 150 × 1500 | 105 × 1000 | 100 × 500 | 45 × 4000 | 10 to 20 length | 0.34 × 0.31 | 90 × 1100 |
| Mobility (cm$^2$/VS) | 0.56 | 7.67 | 0.45 | 0.1 to 0.006 | 340 | 9.7 | 1.9 |
| SS (mV/Dec) | 2300 | | 1000 to 1500 | | | 250 | 170 |
| V$_{th}$ (V) | −0.82 | | | | | 0.5 | −0.16 |
| Operation Voltage (V) | −4 to 4 | <10 | 10 to −30 | 100 to −100 | −60 to 40 | 2 | 20 |
| min bending radius (mm) | 0.75 | 3 | 2 | 5 | 10 | | 6.25 |
| I$_{ON}$/I$_{OFF}$ | 10$^5$ | 10$^4$ to 10$^5$ | | | | 10$^5$ | 10$^8$ |
| Yield | | | | | | | |
| Bending cycles | 10$^4$ | | | | | | |

**Table 3.** Summary of demonstrated hybrid flexible devices over the past half-decade.

| Reference | [178] | [179] | [180] | [181] | [182] | [183] | [184] | [28] | [185] |
|---|---|---|---|---|---|---|---|---|---|
| Year | 2013 | 2014 | 2013 | 2014 | 2014 | 2014 | 2014 | 2014 | 2011 |
| Structure | Parylene/Au gate/Parylene dielectric/pentacene channel/Au source and drain | Au source and drain/(P(NDI2OD)T2 or N2200) n-type or (pBTTT) p-type/PMMA dielectric/Au gate | PVA/PMMA/Au source and drain/MoS$_2$/Al$_2$O$_3$/Cu gate | PEN/Buffer/Al gate/Al$_2$O$_3$/a-IGZO/Mo-Al-Mo source and drain/PR protection | PI/Ni gate/PVP/Al$_2$O$_3$/a-IGZO/Ni source and drain | PDMS/CuPc NW/metal source and drain/Si$_3$N$_4$/Metal Gate | PI/Cu and MoTi Source and drain/organic semiconductor OSC active/organic gate insulator OGI/Cu gate | PI/ Inorganic Si FinFET | PI/IGZO TFT/PI |
| Approach | Hybrid (Transfer) | Hybrid (Transfer) | Hybrid (Transfer) | Hybrid (Deposition at Low Temperature) | Hybrid (Deposition at Low Temperature) | Hybrid (Transfer) | Hybrid (Transfer) | Hybrid (Transfer) | Hybrid (Transfer + encapsulation) |
| Dimensions (μm) | 2 to 10 gate length | 20 × 1000 | 4.3 × 10 | 100 × 100 | 60 × 800 | 10 × 200 | 6 × (14–160) | 0.25 × 3.6 | 115 × 280 |
| Mobility (cm$^2$/VS) | 0.0013 to 0.00022 | 0.1 to 0.3 | 19 | 11.2 | 5.3 to 8.39 | 2 | 0.34 | 141.53 (N) to 13.22 (P) | 13.7 |
| SS (mV/Dec) | ~5000 | | 250 | 270 | 520 to 960 | | | 80 (N) to 70 (P) | |
| V$_{th}$ (V) | −4 to 9.5 | | −2.12 | 0.5 | 7.03 to 9.8 | | | 0.345 (N) to 0.713 (P) | 0.154 |
| Operation Voltage (V) | −32 | 20 to 50 | −5 to 2 | 10 to 10 | 20 to −10 | −10 | 30 to −30 | −1.5 to 1.5 | 5 |
| Switching time (ms) | | | | | | | | | |
| min bending radius (mm) | 0.4 to 0.8 | 1 | 5 | 10 | 10 | 3 | 4 | 5 | 0.125 |
| I$_{ON}$/I$_{OFF}$ | 10$^4$ | 10$^5$ | 10$^6$ | 10$^9$ | 1.4 × 10$^5$ to 3.5 × 10$^6$ | 10$^4$ | 10$^7$ | 10$^{4.6}$ (N) to 10$^{4.78}$ (P) | |
| Yield | | 90% | | | | | | 60% | |
| Bending cycles | | | 10 | 10$^6$ | 10$^5$ | 450 | 10$^5$ | | |

**Table 3.** *Cont*.

| Reference | [186] | [187] | [188] | [189] | [190] | [191] | [192] | [193] |
|---|---|---|---|---|---|---|---|---|
| Year | 2012 | 2013 | 2014 | 2014 | 2015 | 2015 | 2015 | 2013 |
| Structure | PI/SiO$_2$-SiN$_x$-SiO$_2$-SiN$_x$-SiO$_2$/Mo gate/ Hybrid organic-inorganic S-ALO/Au Source and Drain/TIPS PEN:PS Channel | PI/IGZO TFT | Kapton Tape/IZO TFT/ | CPI/IGZO TFT | PI/ZnO TFT | PI/PVP/Al gate/PVP/ZrO2:B/In$_2$O$_3$/Al | PVP/AlO$_x$ /aZITO/ AryLite | Kapton/PI/Ti/Pd/Al$_2$O$_3$ or HfO$_2$/MOS$_2$/Ti/Au |
| Approach | Hybrid (Detachment from Glass Substrate) | Hybrid (Deposition at Low Temperature) | Hybrid (Deposition at Low Temperature) | Hybrid (Deposition at Low Temperature) | Hybrid (Deposition at Low Temperature + Stripping off Si | Hybrid (Deposition at Low Temperature) | Hybrid (Deposition at Low Temperature) | Hybrid (Deposition at Low temperature) |
| Dimensions (μm) | 4 × 20 to 6000 | 6 × 50 | 50 × 1000 | L 20 × W 10 | 20 × 200 | | 50 × 1000 | 1 × 3 |
| Mobility (cm$^2$/VS) | 0.61 | 18 | 6 to 11.2 | 12.7 | 12 | 0.42 | 10.9 | 30 |
| SS (mV/Dec) | 200 | | 2000 to 4600 | 160 | | | | 82 |
| V$_{th}$ (V) | −0.27 | 0.9 | 18.8 to 31.3 | −1.7 | | 8.07 | 0.7 | −2 |
| Operation Voltage (V) | 0.5 to 2 | 5 | | −20 to 20 | 8 | −40 to 40 | 3 | 3 |
| Switching time (ms) | | | | <0.8 × 10$^{-3}$ | | | | |
| min bending radius (mm) | 2 | 5 | 10 | 2 | 3.3 | 5 | 10 | 1 |
| I$_{ON}$/I$_{OFF}$ | 10$^7$ | | 3.8 × 10$^4$ to 1.5 × 10$^6$ | | 10$^8$ | 3.69 × 10$^5$ | 10$^5$ | 10$^7$ |
| Yield | | | | | | | | |
| Bending cycles | 5000 | | 1000 | 1000 | 5 × 10$^4$ | 10$^4$ | 100 | |

*4.2. Hybrid Transistors*

Table 3 highlights the demonstrated hybrid flexible devices. Evidently, device flexibility was lower than for those with all-organic materials such that more than 80% of the devices had a minimum bending radius of ~2 mm or above. This result is expected because the process is more complex and involves inorganic and inflexible materials. Nevertheless, an exceptional dynamic stress stability of up to one million bending cycles was demonstrated for a 10-mm bending radius [181]. The hybrid approach showed mild improvements over the all-organic transistors with mobility values often above the 10 $cm^2$/V.s threshold; although, one reached 141 $cm^2$/V.s [28] with lower SS and operating voltages. Also minimum features of a few microns were demonstrated in many cases, an order of magnitude improvement from the tens of microns features for all organic transistors, with a few exceptions of submicron-scale gate length devices [28,37,74]. Also an order of magnitude improvement in the best $I_{ON}/I_{OFF}$ ratio [181], and 90% yield have been reported [179].

*4.3. Inorganic Transistors on Flexible Silicon*

Table 4 highlights works on flexible inorganic FETs. Two types of inorganic flexible CMOS transistors on Si have been reported: traditional planar metal oxide semiconductor FET (MOSFET) [125,194,195] and 3-dimentional out-of-plane architectural Fin-FET [45]. The electrical and mechanical reliability aspects of the flexed transistor gate stacks, made up of a high-dielectric constant ($\kappa$); ALD $Al_2O_3$, has also been reported and shows a lifetime degradation from electrical stress when the stacks are flexed [196]. The degradation is attributed to increased interfacial charges, leading to a 20% decrease in the safe operational voltage, which would satisfy the ten-year projected lifetime industry standard. A further mechanical reliability assessment was done by observing the effect of mechanical stress on the breakdown voltages of the devices [197]. The results showed that (i) the breakdown voltage increased with more severe bending (lower bending radius/higher strain and stress); (ii) constant mechanical stress might have the same effect as constant electrical stress, and, most notably; (iii) the most severe degradation occurred in devices under dynamically varying mechanical stress (limited number of bending cycles ~100). Otherwise, functionality was reported to pass 200 bending cycles for planar MOSFETs [195], as reliability studies impose harder stress conditions than a device would normally experience for accelerated tests. The minimum bending radii increased considerably to 0.5 mm [45], while the SS, operation voltages, and most importantly, minimum features (tens of the nanometer scale [195]) scaled down. Peak reported mobility of 252 $cm^2$/V.s was reported by Zhai *et al.*, in 2012 [125].

To conclude, the CMOS based flexible transistors are promising candidates for future flexible IoT devices because of their monolithic integration ability, superior electronic properties inherent from bulk form, and almost uncompromised reliability. However, the assessed degradation in performance and safe operation voltages should be accounted for when designing flexible electronic systems utilizing the flexible devices. Another important milestone to achieve fully flexible electronic systems and devices for IoT applications is demonstrating suitable flexible NVMs that can be co-integrated with the flexible transistors for memory storage, without compromising integration density, system speed, and reliability.

Table 4. Works on flexible inorganic field effect transistors (FETs) between January 2010 and May 2015.

| Reference | [194] | [45] | [125] | [195] |
|---|---|---|---|---|
| Year | 2013 | 2014 | 2012 | 2012 |
| Structure | Si/NiSi/Al source and drain/Al$_2$O$_3$ dielectric/TaN-Al gate | Inorganic FinFETs on Thinned Si | Inorganic Planar MOSFETs on Exfoliated Si | Inorganic Planar MOSFETs on Thinned Si |
| Approach | Inorganic (Etch-protect-Release) | Inorganic (Soft Back Etch) | Inorganic (Exfoliation) | Inorganic (Spalling) |
| Dimensions (μm) | 8 length × 5 width | 0.25 (P) to 1 (N) × 3.6 | 0.15 to 1 length | 0.03 length |
| Mobility (cm$^2$/VS) | 43 | 102 (P) | 252 (N) to 51 (P) | |
| SS (mV/Dec) | 80 | 150 (N) to 63 (P) | 81 (P) to 72 (N) | |
| $V_{th}$ (V) | −0.44 | 0.36 (N) to −0.556 (P) | | 0.25 |
| Operation Voltage (V) | 1 to −2 | −1.25 to 1.25 | −1 to 1 | 0.6 |
| Switching time (ms) | | | | <16 × 10$^9$ |
| min bending radius (mm) | | 0.5 | | 6.3 |
| $I_{ON}/I_{OFF}$ | 10$^4$ | 10$^5$ | 10$^6$ (P) | 10$^5$ |
| Yield | | 75% | | |
| Bending cycles | 5 | | | 200 |

## 5. Flexible NVM Technologies

The most important figures of merit for assessing memory technologies are listed below:

a) Form Factor (F$^2$): Although form factor usually refers to the physical lateral (length and width) and vertical (height) dimensions for memory module millimeters, at the device level, the form factor is defined as the lateral area of a single memory cell (1 bit) divided by the square of the smallest feature (technology node) and has no units. For example, a memory cell that is 1 × 0.5 μm$^2$ built at the 0.25-μm node would have a form factor of 8F$^2$.
b) Density: The number of memory bits that fit per unit area.
c) Cost ($/bit): The total cost to make memory modules divided by the number of integrated memory bits.
d) Endurance: The number of write/erase cycles a memory cell undergoes before its performance degrades significantly.

e)  Retention: The retaining ability of a memory cell to store uncompromised information over time.

f)  Operation voltage: The maximum voltage required for a write/erase operation of a memory bit.

g)  Speed: The amount of time the memory cell needs to switch between different memory states ('0' or '1').

h)  Memory window: Measures the distinguishability of the different memory states. Voltage-sensitive memory is proportional to the voltage shift, while current sensitive memory is proportional to the current ratio for different states.

The lower form factor means that memory cells can be arranged efficiently, resulting in reasonable dimensions for the memory array. The higher the integration density, the more bits can be integrated on a specific real estate substrate area (*i.e.*, smaller dies) and the lower the overall bit/cost. Higher endurance is desirable for multiple writing and erasing of data from memory cells and a retention of ten years is the benchmark industry standard for NVMs. Lower operation voltage and high speeds translate into lower power consumption because the switching time when maximum power is drawn is lower as well as the supplied voltage, a sizeable concern for portable and battery operated systems. Furthermore, fast switching means that the NVM can support a faster execution of instructions during processing and computations. The larger the memory window, the more lenient the requirements on the sense circuitry needed to differentiate whether the stored bit is a '0' or a '1'.

*5.1. Flexible ReRAM*

ReRAM or memristors are passive circuit elements that can have two resistance values: high and low resistance states. They act as NVM elements by sensing their resistance during a read cycles or switching their resistance state during write cycles. Memristors were conceptualized by Leon Chua in 1971 [146] and first experimentally demonstrated in 2008 by a team led by R. Stanley Williams at Hewlett Packard labs in 2008 [145]. Since then, memristors have grabbed the attention of the scientific community due to their simplistic structure, fast switching, and possible applications for neuromorphic computations. In 2010, S. Jo *et al.* experimentally demonstrated that CMOS neurons and memristor synapses in a crossbar configuration can support synaptic functions [144]. Extensive research on memristors led to the report of 10 nm × 10 nm ReRAMs at the in International Electron Device Meeting (IEDM) in 2011 [143]. These dimensions were targeted for flash by 2020, based on the ITRS report of 2011 [198]. Memristors are usually integrated in a very dense architecture of cross bars. Furthermore, there are some proposed techniques for using memristor single cells in memory arrays without access/select transistors and for avoiding sneak paths by using a multiple reading sequence [199].

Besides the common perks of flexible electronics, and hence memories, ranging from portable, lightweight, stylish designs, and conformal ability consumer electronics to biomedical applications, flexible memristors would not only support these functionalities but also provide a feasible route for mimicking our brain's cortex structure. Key works on flexible memristors can be classified into four main categories: organic memristors on organic substrates [81,86,87,200–205], inorganic resistive memories on silicon transferred to organic substrates [27,206], inorganic memristors deposited at low temperatures on plastic organic substrates [78,207–217], and inorganic memristors on flexed silicon using the etch-protect-release approach [218]. In addition, interesting work using inorganic flexible substrate (Al foil) with organic cellulose nanofiber paper enabled achieving the lowest reported bending radius for ReRAM (0.35 mm) and lowest operating voltage (±0.5 V) [219]. Table 5 summarizes the key works on flexible ReRAM over the past five years.

*5.2. Flexible FeRAM*

In general, FeRAMs have superior endurance and low variability which represent critical challenges for state-of-the-art redox memristive memories [220]. The two common FeRAM memory architectures using access transistors and ferroelectric capacitors are the 1T-1C, which consists of one transistor and one capacitor per memory cell and 2T-2C, which consists of two transistors and two capacitors for each memory cell [221]. Ferroelectric materials have bi-stable polarization states that can provide useful information for NVM applications, with one state corresponding to a '1' and the other to a '0'. Hence, they are used in simple metal/ferroelectric/metal structures to make highly scalable ferroelectric capacitors, suitable for ultra-high density integration. Rigid ferroelectric random access memories (FeRAM) have already made a great leap by their introduction to the market; hence, it is a relatively mature technology compared to other emerging NVM technologies. FeRAMs are commercially available in Texas Instruments' microprocessors and Fujitsu's RF tags [222–224]. The commonly used ferroelectric material in FeRAM is lead zirconium titanate ($Pb_{1.1}Zr_{0.48}Ti_{0.52}O_3$—PZT) due to its high switching speed [225,226], low cost per bit ($/bit), and low operation voltage [227,228]. Hence, the best properties for flexible NVM FeRAM are reported for structures incorporating PZT as the ferroelectric material, as evident from Table 6. However, PZT-based flexible FeRAM was not achievable on organic platforms because of PZT's high crystallization temperature (>600 °C), well above the melting temperature of most polymeric organic substrates. Hence, PZT FeRAMs have been demonstrated only on flexible silicon and platinum foil [229,230] or built on silicon and then transferred to polymeric flexible substrates [84,231]. Nevertheless, the highest level of mechanical durability (20,000 bending cycles) has been reported for hybrid FeRAMs with inorganic devices transferred to a polymeric substrate [232,233]. Further studies have been conducted on PZT-flexible FeRAM under the combined effect of high temperature (225 °C) and bending condition (1.25 cm bending radius—corresponding to 0.16% nominal strain and ~260 MPa pressure) [234]. These researchers showed evidence of a trend for degradation at higher temperatures but also that temperatures did not affect the retention or the endurance/fatigue properties; however, the memory window (defined as the ratio between switching and nonswitching currents) was significantly reduced at higher temperatures. Interestingly, by capitalizing on standard industry processing techniques, the flexible PZT ferroelectric NVM has the potential to achieve properties similar to those of the best available bulk devices; these are summarized in Table 7.

**Table 5.** Summary of the key works on flexible resistive random access memory (ReRAM) over the past five years.

| Reference | [205] | [217] | [212] | [204] | [206] | [211] | [210] |
|---|---|---|---|---|---|---|---|
| Year | 2010 | 2012 | 2013 | 2012 | 2014 | 2015 | 2012 |
| Memory Type | (1R) | (1R) | (1T) | (1R) | (1D1R) | (1R) | (1R) |
| Flexible Final Structure | PET/Ti/Au/Al/PI:PCBM/Al | PEN/Au/Ag$_2$Se/Ag | PI/SiO$_2$/Ti source and drain/a-IGZO/Ni gate | PEN/Al bottom/coPI layer/ Al top | PI/Si-p-n types diode/Cu/CuOx/Al | PET/ITO/WO$_3$.H$_2$O nanosheets/Cu | Glass/graphene/SiOx/Graphene |
| Approach | All organic | Hybrid (low temperature deposition) | Hybrid (Deposition at Low Temperature) | All organic | Hybrid (Transfer + Low Temperature Deposition) | Hybrid (Low Temperature Deposition) | Hybrid (Low Temperature Deposition) |
| Operating Voltage (V) | 4.5 | −2 | −0.5 to +2 | −3 | +2 and +5 | −1.4 to +1 | 0 to +14 |
| Form Factor (F$^2$) | -- | -- | 2 | -- | 11.1 | -- | -- |
| Memory Window (V) | -- | -- | 1 | -- | 1 | -- | -- |
| Speed (ns) | | -- | 500 × 10$^3$ | 1000 | 5 × 10$^6$ | -- | 50 |
| Endurance (cycles) | 50 | 10$^4$ | 10$^6$ | 10$^4$ | 100 | 5000 | 400 |
| Retention (s) | 10$^4$ | 10$^5$ | 10$^4$ | 10$^4$ | 10$^5$ | 10$^5$ | 5 × 10$^4$ |
| Operating temperature (°C) | 25 | 200 | 85 | 25 | 25 | 25 | 25 |
| Bending Radius (mm) | 9 | 16 | 10 | 5 | 10 | 8 | 6 |
| Bending Cycles | 140 | 100 | 1000 | 1000 | 1000 | 2000 | 300 |
| Yield | -- | -- | -- | 95%–99% | 85%–90% | -- | 70% |
| Cell Dimensions (μm) | -- | -- | 10 × 20 channel | 200 × 200 / 600 × 600 | 150 × 300 channel / 500 cell | | 100 diameter |

Table 5. *Cont*.

| Reference | [87] | [27] | [203] | [86] | [209] | [208] |
|---|---|---|---|---|---|---|
| **Year** | 2010 | 2011 | 2010 | 2010 | 2012 | 2014 |
| **Memory Type** | (1R) | (1T1R) | (1R) | (1R) | (1R) | (1R) |
| **Flexible Final Structure** | PES/Al/graphene-O/Al | Plastic/transferred Si channel material/Au source and drain-> on drainAl/TiO$_2$/Al + Au for WL,BL, and SL | PET/ITO/PMMA/ graphene/PMMA/Al | PET/ITO/ graphene-O/Al | PES/Al/ZrO$_2$/Al | Kapton/Cu/CuO$_x$/Ag |
| **Approach** | All organic | Hybrid (Transfer) | All organic | All organic | Hybrid (Low Temperature Deposition) | Hybrid (Low Temperature Deposition) |
| **Operating Voltage (V)** | −4 | −4 to +10 | −5 | −2 to +3 | −0.5 to 2.8 | −1 to +1 |
| **Form Factor (F$^2$)** | -- | 20 | -- | -- | -- | -- |
| **Memory Window (V)** | -- | 4 | -- | -- | -- | -- |
| **Speed (ns)** | -- | -- | 10$^6$ | -- | -- | -- |
| **Endurance (cycles)** | 100 | 100 | 1.5 × 10$^5$ | 100 | 778 | 100 |
| **Retention (s)** | 10$^5$ | 10$^4$ | 10$^5$ | 5 × 10$^6$ | 10$^5$ | 120 × 10$^4$ |
| **Operating temperature (°C)** | 25 | 25 | 25 | 25 | 25 | −18 to 82 |
| **Bending Radius (mm)** | 7 | 5 | 10 | 4 | -- | 5 |
| **Bending Cycles** | 1000 | 1000 | -- | 1000 | -- | 1000 |
| **Yield** | 80% | 60% | -- | -- | -- | -- |
| **Cell Dimensions (μm)** | 50 × 50 | 10 × 200 | 17.74 to 26.71 diameter | -- | -- | 20 × 20 |

**Table 5.** *Cont*.

| Reference | [207] | [219] | [78] | [218] | [81] | [202] | [201] |
|---|---|---|---|---|---|---|---|
| Year | 2012 | 2014 | 2015 | 2014 | 2012 | 2014 | 2015 |
| Memory Type | (1R) | (1R) | (1R) | (1R) | (1D-1R) | (1R) | (1R) |
| Flexible Final Structure | PES/Cu/TiO$_2$/Cu | Al foil/Ag-CNP(cellulose nanofiber paper)/Ag | PET/Au/Black Phosphorous Quant Dots (BPQD)-PVP/Ag | Si/SiO$_2$/Al/TaN/Al$_2$O$_3$/TaN/Al | PI/Al/B-CNT and N-CNT in polystyrene/Al | PET/rGo/g-C$_3$N$_4$-NSs/rGo | Au/HKUST-1/Au/PET |
| Approach | Hybrid (Deposition at Low Temperature) | Hybrid (Inorganic Flexible Substrate + Organic Device) | Hybrid (Deposition at Low Temperature) | Inorganic | All Organic | All Organic | All Organic |
| Operating Voltage (V) | 0 to 1.5 | −0.5 | −1.2 to +2.8 | −11 to +11 | 3 | 4.87 | 0.78 |
| Form Factor (F$^2$) | -- | -- | -- | -- | -- | -- | -- |
| Memory Window (V) | -- | -- | -- | -- | -- | -- | -- |
| Speed (ns) | -- | -- | -- | -- | -- | -- | -- |
| Endurance (cycles) | -- | 100 | -- | -- | 100 | 50 | 10$^6$ |
| Retention (s) | -- | 10$^5$ | 1.1 × 10$^3$ | -- | 10$^5$ | 5000 | 10$^4$ |
| Operating temperature (°C) | -- | 25 | 25 | 25 | 25 | 25 | −70 to +70 |
| Bending Radius (mm) | 10 | 0.35 | -- | 1 | 10 | 8 | 3.2 |
| Bending Cycles | 100 | 1000 | -- | -- | 500 | 1000 | 160 |
| Yield | -- | -- | -- | -- | -- | -- | -- |
| Cell Dimensions (μm) | -- | 50 × 50 <br> 500 × 500 | 500 × 500 | 100 × 100 <br> 250 × 250 | -- | 1000 × 3000 | 100 diameter |

**Table 6.** Summary of the key works on flexible ferroelectric random access memories (FeRAM) over the past five years.

| Reference | [84] | [231] | [235] | [232] | [233] | [236] | [229] |
|---|---|---|---|---|---|---|---|
| Year | 2013 | 2010 | 2012 | 2011 | 2011 | 2013 | 2015 |
| Memory Type | Ferroelectric (1T) | Ferroelectric (1C) | Ferroelectric (1C) | Ferroelectric (1T) | Ferroelectric (1T) | Ferroelectric (1C) | Ferroelectric (1C) |
| Flexible Final Structure | PI/Su-8/SiO$_2$/Pt gate electrode/ PZT/graphene channel/ Cr-Au for source and drain electrodes | Plastic/Cr-Au/Ti-Pt/PZT/Pt/Cr/Au | Al foil/ PVD-TrFE/ Au | PEN/Ti-Au-Ti soucre and drain/Al$_2$O$_3$/ZnO channel/Al$_2$O$_3$ interface dielectric/PVDF-TrFE ferroelectric/Au | Au/poly(vinylidene fluoride-trifluoroethylene)/Al$_2$O$_3$/ZnO/Ti/Au/Ti/ poly(ethylene naphthalate) | ULTEM 1000B/PEDOT:PSS/P(VDF-TrFE)/PEDOT:PSS | Si/SiO$_2$/Ti-Pt/PZT/Pt |
| Approach | Hybrid (Transfer) | Hybrid (Transfer) | Hybrid (Transfer) | Hybrid (Low Temperature Deposition) | Hybrid (Low Temperature Deposition) | All organic (ink-jet printing) | Inorganic |
| Operating Voltage (V) | −11 | −3 | −12 | −14 to +12 | −10 to +8 | −30 | −15 |
| Form Factor (F$^2$) | 8 | -- | -- | 2 | 2 | -- | -- |
| Memory Window (V) | 6 | -- | -- | 7.8 | 3.4 | -- | -- |
| Speed (ns) | -- | -- | -- | $1 \times 10^9$ | $2 \times 10^9$ | $50 \times 10^6$ | 500 |
| Endurance (cycles) | 1000 | -- | -- | -- | -- | 45% polarization after $10^5$ | $10^9$ |
| Retention (s) | 200 | -- | -- | $1.5 \times 10^4$ | $<10^4$ | -- | $10^5$ |
| Operating temperature (°C) | 25 | 25 | 25 | 25 | 25 | 25 | 200 |
| Bending Radius (mm) | 9 | 8 | 6 | 9.7 | 9.7 | -- | 5 |
| Bending Cycles | 200 | -- | 500 | $2 \times 10^4$ | $2 \times 10^4$ | -- | 1000 |
| Yield | -- | -- | -- | -- | -- | -- | 95% |
| Cell Dimensions (μm) | 10 × 80 channel | 100 × 400 | 180 diameter | 20 × 40 channel | 20 × 40 channel | 60 × 60 | 100 × 100 <br> 250 × 250 |

**Table 6.** *Cont.*

| Reference | [237] | [238] | [239] | [240] | [230] | [241] | [242] | [243] |
|---|---|---|---|---|---|---|---|---|
| **Year** | 2011 | 2011 | 2012 | 2012 | 2012 | 2015 | 2014 | 2013 |
| **Memory Type** | Ferroelectric (1C) | Ferroelectric (1C) | Ferroelectric (1T) | Ferroelectric (1C) | Ferroelectric (1C) | Ferroelectric (1R) | Ferroelectric (1T) | Ferroelectric (1T) |
| **Flexible Final Structure** | PEN/PEDOT:PSS/P(VDF-TrFE)/PEDOT:PSS | PEN/Au/P(VDF-TrFE)/Au | Bank Note/PDMS/PEDOT:PSS bottom electrode/ P(VDF-TrFE) ferroelectric/ Pentacene channel/ Au soucre and drain | Ag/BaTiO$_3$/PVDF-TrFE nanocomposites/Ag | Pt/PZT (200 nm)/SRO (30 nm)/Pt (200 nm) foil | PET/Ag ink/PVDF-TrFE/Ag ink | PDMS/Au source and drain/F8T2 organic semiconductor/PVDF-TrFE ferroelectric/ Al gate | PI/PVP/Au source and Drain/Polymer blend channel/PVDF-TrFE ferroelectric/Al gate |
| **Approach** | All organic | All organic | All organic | Hybrid (inorganic/organic composite substrate and device) | Inorganic | All Organic | All Organic | All Organic |
| **Operating Voltage (V)** | −10 | −30 to +30 | −15 to +15 | −3.3 | −4 | 23 | −20 to +20 | −80 to +80 |
| **Form Factor (F$^2$)** | -- | -- | 16.67 | -- | -- | -- | 3 | 1 |
| **Memory Window (V)** | -- | -- | 8 | -- | -- | -- | 11 | 35 |
| **Speed (ns)** | 10$^6$ | 10 × 10$^3$ | -- | -- | -- | 10$^8$ | 10$^8$ | -- |
| **Endurance (cycles)** | 10$^6$ has 80% Pr | -- | 10$^5$ | -- | 10$^7$ | -- | 6 × 10$^4$ | 100 |
| **Retention (s)** | -- | -- | 10$^4$ | -- | -- | -- | 2000 | 10$^4$ |
| **Operating temperature (°C)** | 25 | 25 | 25 | 25 | 25 | 25 | 25 | 25 |
| **Bending Radius (mm)** | -- | 6.5 | -- | -- | -- | -- | 6 | -- |
| **Bending Cycles** | -- | -- | -- | -- | -- | -- | -- | -- |
| **Yield** | -- | -- | 90% | -- | -- | -- | -- | -- |
| **Cell Dimensions (μm)** | -- | 25 × 25<br>50 × 50 | 60 × 1000 channel | -- | 100 diameter | 30 × 30 | 20 × 60 | 20 × 20 |

**Table 7.** Summary of best reported values for PZT-based ferroelectric nonvolatile memory (NVM).

| Property | Best Reported Value for PZT-based FeRAM |
|---|---|
| Switching Speed | Pico seconds regime for material switching [225,226] and 70 ns for actual arrays due to bit/word line capacitances [244] |
| Ferroelectric Capacitor's Lateral Dimensions | 0.1 µm$^2$ [245] |
| Switching Energy | 400 fJ/bit [246] |
| Retention | >10 years @ 85 °C [247], experimentally three days (2.5 × 10^5) was demonstrated for FeRAM (using SrBi$_2$Ta$_2$O$_9$(SBT)) [248,249] |
| Technology Node (CMOS Logic) | 130 nm [247] |
| Operation Voltage | 1.5 Volts [246] |
| Read/Write Cycles | >10$^{15}$ [250,251] |

### 5.3. Flexible PCRAM

Phase change materials exhibit high resistance in the highly disordered amorphous phase and low resistance in the highly ordered crystalline phase, where each resistance state corresponds to a '0' or a '1'. The transition to the amorphous phase requires applying high temperatures above the melting point of the phase change material for a brief time, while the transition to the crystalline phase requires lower temperatures for a longer duration to supply the required energy for re-organizing the material structure. PCRAM is generally characterized by high switching transition speeds. In 2010, Hong *et al.* reported a flexible version of the Ge$_2$Sb$_2$Te$_5$-based PCRAM on polyimide that required a 30 ns pulse to switch [252]. Another advantage of PCRAM is its highly localized regions of phase change that enables ultra-high integration densities. In 2011, Hong *et al.* also reported phase-change nano-pillar devices with the potential of reaching up to tera bit/squared inch densities on flexible substrates [253] and the following year, Yoon *et al.* demonstrated a 176 Gbit/square inch PCRAM [254], the highest reported density on a flexible substrate. The highest reported bending cycles endurance (1000 bending cycles) and yield (66%) for flexible PCRAM was reported by Mun *et al.*, in 2015 [255]. Table 8 summarizes the work on flexible PCRAM undertaken over the past five years.

**Table 8.** Summary of the key works on flexible PCRAM over the past five years.

| Reference | [252] | [253] | [254] | [255] |
|---|---|---|---|---|
| Year | 2010 | 2011 | 2012 | 2015 |
| Memory Type | PCRAM (1R) | PCRAM (1R) | PCRAM (1R) | PCRAM (1D-1R) |
| Flexible Final Structure | PI/TiN/GST/Pt probe tip | PET or PI or Stainless steel/ TiW bottom electrode/GST ($Ge_2Sb_2Te_5$)/Cr top electrode | GFR Hybrimer Film/TiN/Pt/TiN/IST/Ps-b-PMMA/Cr | BCP/Transferred Si Diode/TiW bottom electrode/GST/$SiO_x$ cylinders/TiN/TiW top electrode |
| Approach | Hybrid (Low Temperature Deposition + Transfer by Hot Embossing and Nano Imprint Lithography -NIL) | Hybrid (Low Temperature Deposition + NIL) | Hybrid (Low Temperature Deposition) | Hybrid (transfer + low temperature deposition) |
| Operating Voltage (V) | 1.8 | 3 | 1.8 | 8.5 |
| Form Factor ($F^2$) | -- | -- | -- | -- |
| Memory Window | -- | -- | -- | -- |
| Speed (ns) | 30 | 200 | 100 | 1000 |
| Endurance (cycles) | -- | -- | -- | 100 |
| Retention (s) | -- | -- | -- | $10^4$ |
| Operating temperature (°C) | 25 | 25 | 25 | 25 |
| Bending Radius (mm) | -- | -- | 2.25 | 10 |
| Bending Cycles | -- | -- | -- | 1000 |
| Yield | -- | -- | -- | 66% |
| Cell Dimensions (μm) | 0.2 to 0.3 diameter | 0.25 diameter | 0.035 diameter | -- |

*5.4. Flexible Flash*

Flash memories are the most mature NVM technology in today's market. Flash memories have the same FET structure with an extra embedded layer for charge trapping/storage for simple nondestructive readout operations [75,256]. The charge-trapping layer is sandwiched between two dielectrics. The first dielectric is a blocking oxide, which is a thick oxide to prevent charge transfer between the control gate and the charge-trapping layer. The second is a tunneling oxide with thicknesses varying between 7 and 10 nm through which charge transfer takes place between the FET channel and the charge-trapping layer. The two aforementioned flash types, the CT-flash and the FG-flash, differ in the nature of the charge-trapping layer used. FG-flash has a floating conductor/gate previously made of doped polysilicon that has recently been shifting to conductive NPs and QDs for charge trapping. CT-flash has an insulating charge-trapping layer. Building on the structural difference, there is also a fundamental difference in the physical operation. The stored charge is removed/erased from a FG-flash memory through Fowler-Nordheim (FN) tunneling between the charge-trapping layer and the FET channel, where a strong electric field is applied at the gate reducing the effective tunneling barrier width [257]; whereas, in hot-hole injection, where an electric field causes a hole with sufficient energy in the channel to tunnel, neutralizing a stored charge in the charge-trapping layer [258], which serves as the erasing mechanism in CT-flash. Both FG-flash and CT-flash are programmed using hot electron injection from the channel to the charge-trapping layer. Compared to FN tunneling, hot carrier injection requires lower electric fields, which allowed for the scaling down of the tunneling oxide thickness without compromising its reliability [70,73,74,256,259]. To overcome the challenges associated with traditional FG-flashes, NPs and nanocrystal FG-flashes have recently been introduced.

Despite the widespread adoption of flash memories in a myriad of applications, the status quo for flexible flash memories does not reflect the perks of the long commercialized rigid flash. This is in part due to the focus on replacing the semiconductor's standard processes and materials in favor of flexible new materials and low-temperature processes. Limited by the organic material choices of the all-organic approach and low-temperature depositions of the hybrid approach to flexible NVMs, current flexible flash memories have reported operation voltages ranging from ±5 to ±90 V [77,260] with minimum channel length dimensions of 2-μm [80]. Nonetheless, good bendability has been achieved up to 5 mm [85,261] for 2000 bending cycles [77], an endurance of 100,000 cycles [76], and a retention ability of $10^6$ s [73]. Table 9 summarizes the work done on flexible flash NVMs since 2010.

To summarize, due to the various approaches towards achieving flexible electronics, the versatility of nonvolatile memory technologies, and relative infancy of emerging NVM technologies, flexible NVM elements face more challenges in order to match their bulk counterparts. For instance, cellulose nanofiber ReRAM on Al foil achieved ±0.5 V operation voltage but integration ability with CMOS technology is a challenge. FeRAM being relatively a mature technology and since its transformation from bulk to flexible form in a CMOS compatible process has been demonstrated, holds potential for suitability for IoT applications. Based on the possible optimizations in Table 7, the flexible FeRAM using the inorganic approach can potentially achieve low power operation, high endurance, 10 years retention, and the ability to be integrated in 130 nm CMOS technology. However, the demonstrating this potential has not been reported. Although a flexible 176 Gb/square inch has been demonstrated, it suffers from low yield. This is a critical challenge when discussing application in IoT and big data analysis where gigantic memory arrays are to be utilized because faulty memory bit provides wrong information and can have a detrimental effect on neighboring memory cells in the array. Flexible flash NVMs are far from state-of-the-art bulk flash properties because most of the research done focused on new material systems and processes.

In the IoT era, where smart electronic devices are able to make decisions autonomously without human intervention for application in fully automated cars that car drive itself at high speeds the integrity of electronic systems and its ability to store and process large pieces of information in a fractions of a second cannot be compromised. CMOS based electronics have demonstrated great integrity, as well as fast and reliable performance for decades through the well-established and heavily invested infrastructures and standard processes. Although the fact that a flexible version of state-of-the-art CMOS transistors has been demonstrated holds promise for flexible electronics, flexible NVMs are still the weaker link in the chain and without NVMs, fully flexible electronic systems cannot be achieved.

## 6. Conclusions and Future Prospects

Exciting progress has been made in flexible electronics research over the past few decades. OLED flexible screens are already available in the market, and numerous novel biomedical and wearable applications using flexible electronics have been proposed. At this stage, expectations for silicon-based electronics are high and the *status quo* is for high performance, fast, low power, compact, and reliable aspects, some of which might not cross the chasm to the flexible arena. This is the core value for which the research field of transfer-free inorganic silicon-based flexible electronics is created. However, this approach is relatively new compared to the ongoing research on flexible organic electronics, where organic materials are used as substrates or device material.

In this review, we have presented a brief overview of flexible electronics research, focusing on NVM components. We listed the mainstream NVM architectures and technologies with a special focus on flexible devices and provided benchmarking tables for transistors and storage devices derived from the three main approaches: (1) all organic; (2) hybrid; and (3) inorganic.

Future prospects indicate that a number of challenges will have to be overcome before flexible ReRAM, FeRAM, PCRAM, MRAM, and flash will be primed for commercialization. Flexible FeRAM has an edge because of its rigid current form that is used as an embedded NVM in microprocessors. Flash has the highest potential owing to the maturity of the technology in the rigid state; furthermore, the transfer of this technology and progress to the flexible arena will definitely speed up the introduction of commercial flexible flash NVM for various applications. On the other hand, ReRAM, PCRAM, and MRAM are still emerging technologies, even in their bulk rigid form. The potential for extreme scaling, fast speeds, and low-power operation of NVMs is attracting the attention of both researchers and industry such that they might catch up with competing mature technology in the flexible arena earlier than expected.

**Table 9.** Summary of the key works on flexible Flash memory over the past five years.

| Reference | [260] | [82] | [80] | [76] | [77] |
|---|---|---|---|---|---|
| Year | 2012 | 2011 | 2012 | 2014 | 2011 |
| Memory Type | Flash + Nano Particles (1T) | Flash (1T) | Flash (1T) | Flash (1T) | Flash (1T) |
| Flexible Final Structure | PET/Ag gate electrode/Al$_2$O$_3$/PMMA with Au NP composite/Pentacene/Au source and drain electrodes | PET/graphene gate electrode/Al$_2$O$_3$/CNT channel/graphene for source and drain electrodes | PES/Si NW/Al$_2$O$_3$/Pt-NP//Al$_2$O$_3$/Al source, drain, and gate | PET/ P(NDI2OD-T2) Channel/ Au source and drain/PVA tunneling dielectric/ Au-NP/ P(VDF−TrFE−CFE) blocking dielectric/ Al gate | PES/ITO gate/PVP blocking/APTES-Au-NP/PVP tunneling/MoO$_3$ buffer/ Pentacene channel/ITO source and drain |
| Approach | All organic | All organic | Hybrid (Transfer) | All organic | All organic |
| Operating Voltage (V) | −5 to +5 | −10 to +10 | −10 to +10 | −6 to +6 | −90 to +90 |
| Form Factor (F$^2$) | 10 | -- | -- | 20 | 10 |
| Memory Window | 2.1 | 10 | 1.85 | 2 | 15 |
| Speed (ns) | -- | 100 | 10$^7$ | 2 × 10$^9$ | 1 × 10$^9$ |
| Endurance (cycles) | 1000 | 500 | 10$^4$ | 10$^5$ | -- |
| Retention (s) | 10$^5$ | 1000 | 10$^4$ | 10$^5$ | 10$^5$ |
| Operating temperature (°C) | 100 | 25 | 25 | 25 | 85 |
| Bending Radius (mm) | 10 | 8 | 16 | 9 | 20 |
| Bending Cycles | 1000 | 1000 | 1000 | 100 | 2000 |
| Yield | -- | -- | -- | -- | -- |
| Cell Dimensions (μm) | 50 × 500 channel | -- | 2 channel length | 100 × 2000 channel | 100 × 1000 channel |

**Table 9.** *Cont.*

| Reference | [262] | [74] | [85] | [73] | [261] |
|---|---|---|---|---|---|
| Year | 2012 | 2010 | 2012 | 2013 | 2015 |
| Memory Type | Flash (1T) | Flash (1T) | Flash (1T) | Flash (1T) | Flash (1T) |
| Flexible Final Structure | PES/Al source and drain/ZnOPDA/AlO$_x$-SAOLs/Zno:Cu/AlO$_x$-SAOLs/Pentacene/ Al gate | PES/Ti-Au gate/ PVP blocking/ APTES-Au NP storage/PVP tunneling/ Pentacene channel/ Au source and drain | PEN/graphene channel/Al$_2$O$_3$/HfO$_x$/Al$_2$O$_3$/ ITO | PET/ITO gate/Al$_2$O$_3$/Au NP charge trapping/Al$_2$O$_3$/PDPP-TBT/Au source and drain | PDMS/PI/Au/Al$_2$O$_3$-SiO$_2$/SWCNT/Au |
| Approach | Hybrid (Low Temperature Deposition) | All organic | Hybrid (Deposition at Low Temperature) | All organic | Hybrid (Transfer) |
| Operating Voltage (V) | −15 to +15 | −90 to +90 | −21 to +23 | −40 to +40 | −25 to +25 |
| Form Factor (F$^2$) | 2 | 10 | 6.67 | 20 | 11 |
| Memory Window | 14.1 | 9.7 | 8.6 | 7.5 | 13.2 |
| Speed (ns) | $1 \times 10^8$ | $1 \times 10^9$ | hypothesized ~14 ns read time and 20 μs/20 ms write/erase time | $10 \times 10^7$ | $10^5$ |
| Endurance (cycles) | -- | 700 | -- | 1000 | $10^4$ |
| Retention (s) | 1000 | $10^5$ | 30% after 10 years | $10^6$ | $10^4$ |
| Operating temperature (°C) | 25 | 25 | degrades at 85 | 25 | 25 |
| Bending Radius (mm) | -- | 20 | 5 | 10 | 5 |
| Bending Cycles | -- | 1000 | 10 | 1000 | 1000 |
| Yield | -- | -- | -- | -- | -- |
| Cell Dimensions (μm) | 50 × 100 channel | 100 × 1000 channel | 30 × 4.5 channel | 50 × 1000 channel | 18 × 200 |

These novel flexible NVM technologies along with the required flexible transistors for array gating and data processing are enabling technologies for the foreseen future where a galactic network of connected devices can autonomously detect, collect, and process surrounding information to make real time decisions fulfilling the IoT futuristic vision. The NVMs will store instructions and data in both power-on and power off states consuming minimal power. Furthermore, to meet the ultra-low power requirements for IoT devices when the collected information does not need real time processing, integrating ultra-high density NVMs will help store large amount of information to be analyzed when the device is externally charging, instead of sending data continuously through an antenna and consuming extra power. This is useful when collecting activity level throughout the day, similar to the study conducted by Hitachi that provided a link between variability of motion and happiness. NVMs will also have to cope with ultra-large scale integration to be able to store the huge amounts of information for big data analysis, in compact ultra-mobile devices, that can be preferably monolithically integrated in state-of-the-art CMOS technologies to expedite its introduction in the consumer electronics market. However, the status quo of flexible NVMs shows that there is still a gap between current bulk devices' properties and their flexible counterparts. Hence, flexible NVMs and flexible electronic systems are still hindered from invading the consumer electronics market and still provide less attractive alternative to replacing bulk devices for future IoT applications. The core challenge is to either invent new technologies that can rival CMOS integration ability and standard processes at a reasonable cost, or to embrace CMOS technology in making flexible devices while obtaining flexibility as an extra feature without affecting critical performance metrics of the devices.

## Acknowledgments

We acknowledge the KAUST OCRF CRG-1 grant CRG-1-2012-HUS-008 to MTG. We also thank Carolyn Unck, Editor, Academic Writing of KAUST for proof reading and editing our manuscript.

## Conflicts of Interest

The authors declare no competing financial interest. We have also diligently secured copyrights for already published materials used in this review paper.